\documentclass[]{article}

\usepackage[latin1]{inputenc}
\usepackage{graphicx}
\usepackage{amssymb,amsfonts,amsmath}
\usepackage{theorem}
\usepackage{hyperref}
\usepackage{graphicx}
\usepackage{multirow}

\let\realleq\leq

\newtheorem{corollary}{Corollary}[section]

\newtheorem{proposition}{Proposition}[section]
{\theorembodyfont{\rmfamily}
\newtheorem{definition}{Definition}[section]
\newtheorem{remark}{Remark}[section]
\newtheorem{example}{Example}[section]
}
\newenvironment{proof}[1][Proof]{\noindent\textbf{#1.} }{\newline \hspace*{\textwidth}\hspace*{-0,4cm} \rule{0.5em}{0.5em} \vspace{0,2cm}}

\begin{document}

\title{\textbf{Intrinsic definitions of ``relative velocity'' in general relativity}}

\author{Vicente J. Bol\'os \\{\small Dpto. Matem\'aticas para la Econom\'{\i}a y la Empresa, Facultad de Econom\'{\i}a,}\\ {\small Universidad de Valencia. Avda. Tarongers s/n. 46022, Valencia,
Spain.}\\ {\small e-mail\textup{: \texttt{vicente.bolos@uv.es}}}}

\date{June 2005 \\
(Revised version: April 2011)}

\maketitle

%\PACS 04.20.Cv \sep 02.40.Hw \MSC 53A35 \sep 53B30 \sep 83C99

\begin{abstract}
Given two observers, we define the ``relative velocity'' of one
observer with respect to the other in four different ways. All
four definitions are given intrinsically, i.e. independently of
any coordinate system. Two of them are given in the framework of
spacelike simultaneity and, analogously, the other two are given
in the framework of observed (lightlike) simultaneity. Properties
and physical interpretations are discussed. Finally, we study
relations between them in special relativity, and we give some
examples in Schwarzschild and Robertson-Walker spacetimes.
\end{abstract}

\section{Introduction}

The need for a strict definition of \textquotedblleft radial
velocity\textquotedblright\ was treated at the General Assembly of
the International Astronomical Union (IAU), held in 2000 (see
\cite{Soff03}, \cite{Lind03}), due to the ambiguity of the classic
concepts in general relativity. As result, they obtained three
different concepts of \textit{radial velocity}: \textit{kinematic}
(which corresponds most closely to the line-of-sight component of
space velocity), \textit{astrometric} (which can be derived from
astrometric observations) and \textit{spectroscopic} (also called
\textit{barycentric}, which can be derived from spectroscopic
measurements). The kinematic and astrometric radial velocities
were defined using a particular reference system, called
Barycentric Celestial Reference System (BCRS). The BCRS is
suitable for accurate modelling of motions and events within the
solar system, but it has not into account the effects produced by
gravitational fields outside the solar system, since it describes
an asymptotically flat metric at large distances from the Sun.
Moreover, from a more theoretical point of view, these concepts
can not be defined in an arbitrary spacetime since they are not
intrinsic, i.e. they only have sense in the framework of the BCRS.
So, in this work we are going to define them intrinsically. In
fact, we obtain in a natural way four intrinsic definitions of
relative velocity (and consequently, radial velocity) of one
observer $\beta '$ with respect to another observer $\beta $,
following the original ideas of the IAU.

This paper has two big parts:
\begin{itemize}

\item The first one is formed by Sections \ref{sec3} and
\ref{sec4}, where all the concepts are defined, trying to make the
paper as self-contained as possible. In Section \ref{sec3}, we
define the \textit{kinematic} and \textit{Fermi} relative
velocities in the framework of spacelike simultaneity (also called
Fermi simultaneity), obtaining some general properties and
interpretations. The kinematic relative velocity generalizes the
usual concept of relative velocity when the two observers $\beta
$, $\beta '$ are at the same event. On the other hand, the Fermi
relative velocity does not generalize this concept, but it is
physically interpreted as the variation of the \textit{relative
position} of $\beta '$ with respect to $\beta $ along the world
line of $\beta $. Analogously, in Section \ref{sec4}, we define
and study the \textit{spectroscopic} and \textit{astrometric}
relative velocities in the framework of observed (lightlike)
simultaneity.

\item In the second one (Sections \ref{sec5} and
\ref{sec6}) we give some relations between these concepts in
special and general relativity. In Section \ref{sec5} we find
general expressions, in special relativity, for the relation
between kinematic and Fermi relative velocities, and between
spectroscopic and astrometric relative velocities. Finally, in
Section \ref{sec6} we show some fundamental examples in
Schwarzschild and Robertson-Walker spacetimes.

\end{itemize}

\section{Preliminaries}

We work in a 4-dimensional lorentzian spacetime manifold $\left(
\mathcal{M},g\right) $, with $c=1$ and $\nabla $ the Levi-Civita
connection, using the Landau-Lifshitz Spacelike Convention (LLSC).
We suppose that $\mathcal{M}$ is a convex normal neighborhood
\cite{Helg62}. Thus, given two events $p$ and $q$ in
$\mathcal{M}$, there exists a unique geodesic joining $p$ and $q$
and there are not caustics. The parallel transport from $p$ to $q$
along this geodesic will be denoted by $\tau _{pq}$. If $\beta
:I\rightarrow \mathcal{M}$ is a curve with $I\subset \mathbb{R}$ a
real interval, we will identify $\beta $ with the image $\beta I$
(that is a subset in $\mathcal{M}$), in order to simplify the
notation. If $u$ is a vector, then $u^{\bot }$ denotes the
orthogonal space of $u$. The projection of a vector $v$ onto
$u^{\bot }$ is the projection parallel to $u$. Moreover, if $x$ is
a spacelike vector, then $\Vert x\Vert $ denotes the modulus of
$x$. Given a pair of vectors $u,v$, we use $g\left( u,v\right) $
instead of $u^{\alpha }v_{\alpha }$. If $X$ is a vector field (typically, vector fields will be denoted by uppercase letters),
$X_{p}$ denotes the unique vector of $X$ in
$T_{p}\mathcal{M}$.

In general, we will say that a timelike world line $\beta $ is an
\textit{observer} (or a \textit{test particle}). Nevertheless, we will say that a
future-pointing timelike unit vector $u$ in $T_{p}\mathcal{M}$ is
an \textit{observer at $p$}, identifying it with its 4-velocity.

The relative velocity of an observer (or a test particle) with respect to another
observer is completely well defined only when these observers are
at the same event: given two observers $u$ and $u^{\prime }$ at
the same event $p$, there exists a unique vector $v\in u^{\bot }$
and a unique positive real number $\gamma $ such that
\begin{equation}
u^{\prime }=\gamma \left( u+v\right) .  \label{f1.uprima}
\end{equation}
As consequences, we have $0\realleq\Vert v\Vert <1$ and $\gamma
:=-g\left( u^{\prime },u\right) =\frac{1}{\sqrt{1-\Vert v\Vert
^{2}}}$. We will say that $v$ is the \textit{relative velocity of
}$u^{\prime }$\textit{\ observed by }$u$, and $\gamma $ is the
\textit{gamma factor} corresponding to the velocity $\Vert v\Vert
$. From (\ref{f1.uprima}), we have
\begin{equation}
v=\frac{1}{-g\left( u^{\prime },u\right) }u^{\prime }-u.
\label{relvel}
\end{equation}
We will extend this definition of relative velocity in two
different ways (\textit{kinematic} and \textit{spectroscopic}) for
observers at different events. Moreover, we will define another
two concepts of relative velocity (\textit{Fermi} and
\textit{astrometric}) that do not extend (\ref{relvel}) in
general, but they have clear physical sense as the variation of
the \textit{relative position}.

A \textit{light ray} is given by a lightlike geodesic $\lambda $
and a future-pointing lightlike vector field $F$ defined in
$\lambda $, tangent to $\lambda $ and parallelly transported along
$\lambda $ (i.e. $\nabla _{F}F=0$), called \textit{frequency} (or \textit{wave})
\textit{vector field of }$\lambda $. Given $p\in \lambda $ and $u$ an
observer at $p$, there exists a unique vector $w\in u^{\bot }$ and
a unique positive real number $\nu $ such that
\begin{equation}
F_{p}=\nu \left( u+w\right) .  \label{f1.efepe}
\end{equation}
As consequences, we have $\Vert w\Vert =1$ and $\nu =-g\left(
F_{p},u\right) $. We will say that $w$ is the \textit{relative
velocity of }$\lambda $ \textit{\ observed by }$u$, and $\nu $ is
the \textit{frequency of }$\lambda $\textit{\ observed by }$u$. In
other words, $\nu $ is the modulus of the projection of $F_{p}$
onto $u^{\perp }$. A \textit{light ray from }$q$\textit{\ to }$p$
is a light ray $\lambda $ such that $q$, $p\in \lambda $ and $\exp
_{q}^{-1}p$ is future-pointing.

\section{\label{sec3}Relative velocity in the framework of spacelike
simultaneity}

The spacelike simultaneity was introduced by E. Fermi (see
\cite{Ferm22}), and it was used to define the \textit{Fermi
coordinates}. So, some concepts given in this section are very
related to the work of Fermi, as the \textit{Fermi surfaces}, the
\textit{Fermi derivative} or the \textit{Fermi distance}. The
original Fermi paper and most of the modern discussions of this
notion (see \cite{Marz94}, \cite{Bini05}) use a coordinate
language (Fermi coordinates). On the other hand, in the present
work we use a coordinate-free notation that allows us to get a
better understanding of the basic concepts of the Fermi work,
studying them from an intrinsic point of view and, in the next
section, extending them to the framework of lightlike
simultaneity.

Let $u$ be an observer at $p\in \mathcal{M}$ and $\Phi
:\mathcal{M}\rightarrow \mathbb{R}$ defined by $\Phi \left(
q\right) :=g\left( \exp _{p}^{-1}q,u\right) $. Then, it is a
submersion and the set $L_{p,u}:=\Phi ^{-1}\left( 0\right) $ is a
regular 3-dimensional submanifold, called \textit{Landau
submanifold of }$\left( p,u\right) $ (see \cite{Oliv80},
\cite{Bolo02}), also known as \textit{Fermi surface}. In other
words, $L_{p,u}=\exp _{p}u^{\bot }$. An event $q$ is in $L_{p,u}$
if and only if $q$ is simultaneous with $p$ in the local inertial
proper system of $u$.

\begin{figure}[tbp]
\begin{center}
\includegraphics[width=0.5\textwidth]{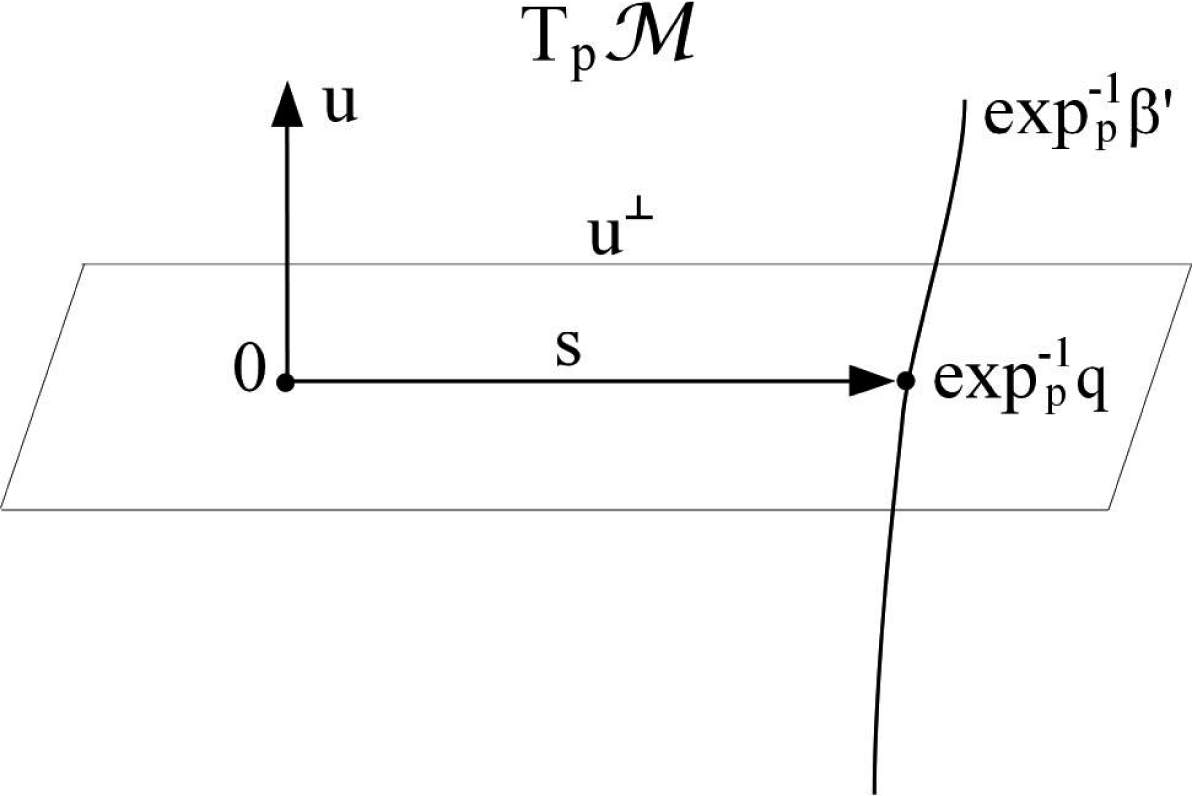}
\end{center}
\caption{Scheme in $T_p \mathcal{M}$ of the relative position $s$
of $q$ with respect to $u$.} \label{figrelpos}
\end{figure}

\begin{definition}
Given $u$ an observer at $p$, and a simultaneous event $q\in
L_{p,u}$, the \textbf{relative position of }$q$\textbf{\ with
respect to }$u$ is $s:=\exp _{p}^{-1}q$ (see Fig.
\ref{figrelpos}).
\end{definition}

We can generalize this definition for two observers $\beta $ and
$\beta ^{\prime }$.

\begin{definition}
\label{relpossp} Let $\beta $, $\beta ^{\prime }$ be two observers
and let $U$ be the 4-velocity of $\beta $. The \textbf{relative
position of }$\beta ^{\prime }$\textbf{\ with respect to }$\beta $
is the vector field $S$ defined on $\beta $ such that $S_{p}$ is
the relative position of $q$ with respect to $U_{p}$, where $p\in
\beta $ and $q$ is the unique event of $\beta ^{\prime }\cap
L_{p,U_{p}}$.
\end{definition}

\subsection{Kinematic relative velocity}

We are going to introduce the concept of ``kinematic relative
velocity" of one observer $u'$ with respect to another observer
$u$ generalizing the concept of relative velocity given by
(\ref{relvel}), when the two observers are at different events.

\begin{figure}[tbp]
\begin{center}
\includegraphics[width=0.4\textwidth]{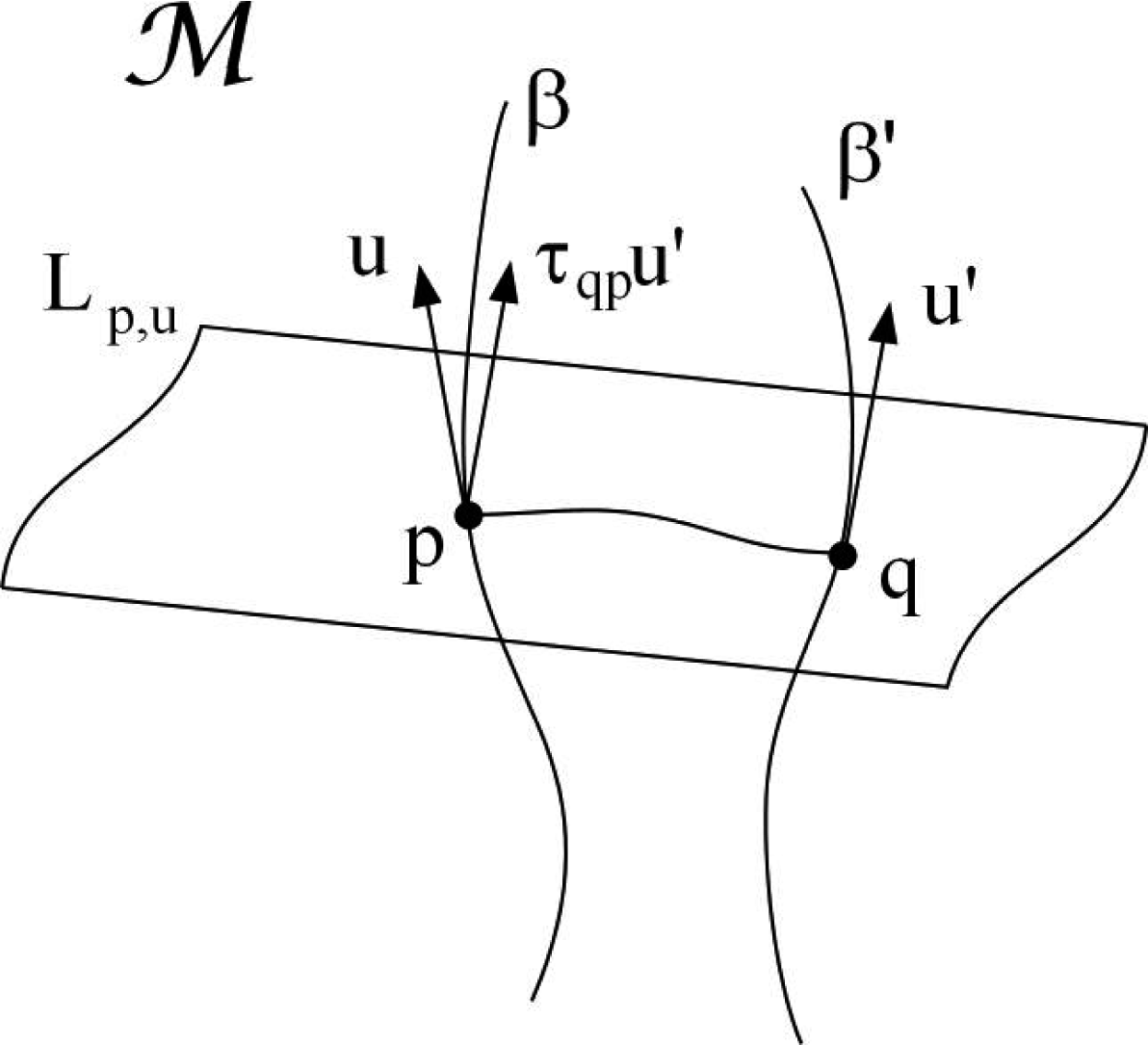}
\end{center}
\caption{Scheme in $\mathcal{M}$ of the elements that involve the
definition of the kinematic relative velocity of $u^{\prime}$ with
respect to $u$.} \label{figvkin}
\end{figure}

\begin{definition}
\label{kinrelvel}Let $u$, $u^{\prime }$ be two observers at $p$,
$q$ respectively such that $q\in L_{p,u}$. The \textbf{kinematic
relative velocity of }$u^{\prime }$\textbf{\ with respect to }$u$
is the unique vector $v_{\mathrm{kin}}\in u^{\bot }$ such that
$\tau _{qp}u^{\prime }=\gamma \left( u+v_{\mathrm{kin}}\right) $,
where $\gamma $ is the gamma factor corresponding to the velocity
$\Vert v_{\mathrm{kin}}\Vert $ (see Fig. \ref{figvkin}). So, it is
given by
\begin{equation}
\label{fkinrelvel} v_{\mathrm{kin}}:=\frac{1}{-g\left( \tau
_{qp}u^{\prime },u\right) }\tau _{qp}u^{\prime }-u.
\end{equation}
Let $s$ be the relative position of $q$ with respect to $p$, the
\textbf{kinematic radial velocity of }$u^{\prime }$\textbf{\ with
respect to }$u$ is the component of $v_{\mathrm{kin}}$ parallel to
$s$, i.e. $v_{\mathrm{kin}}^{\mathrm{rad}}:=g\left(
v_{\mathrm{kin}},\frac{s}{\left\Vert s\right\Vert }\right)
\frac{s}{\left\Vert s\right\Vert }$. If $s=0$ (i.e. $p=q$) then
$v_{\mathrm{kin}}^{\mathrm{rad}}:=v_{\mathrm{kin}}$. On the other
hand, the \textbf{kinematic tangential velocity of }$u^{\prime
}$\textbf{\ with respect to }$u$ is the component of
$v_{\mathrm{kin}}$ orthogonal to $s$, i.e.
$v_{\mathrm{kin}}^{\mathrm{tng}}:=v_{\mathrm{kin}}-v_{\mathrm{kin}}^{\mathrm{rad}}$.
\end{definition}

So, the kinematic relative velocity of $u^{\prime }$ with respect
to $u$ is the relative velocity of $\tau _{qp}u^{\prime }$
observed by $u$, in the sense of expression (\ref{relvel}). Note
that $\Vert v_{\mathrm{kin}}\Vert <1$, since the parallel
transported observer $\tau_{qp}u'$ defines an observer at $p$.

We can generalize these definitions for two observers $\beta $ and
$\beta ^{\prime }$.

\begin{definition}
\label{defkinrelvel}Let $\beta $, $\beta ^{\prime }$ be two
observers, and let $U$, $U^{\prime }$ be the $4$-velocities of
$\beta $, $\beta ^{\prime }$ respectively. The \textbf{kinematic
relative velocity of }$\beta ^{\prime }$\textbf{\ with respect to
}$\beta $ is the vector field $V_{\mathrm{kin}}$ defined on $\beta
$ such that $V_{\mathrm{kin}~p}$ is the kinematic relative
velocity of $U_{q}^{\prime }$ observed by $U_{p}$ (in the sense of
Definition \ref{kinrelvel}), where $p\in \beta $ and $q$ is the
unique event of $\beta ^{\prime }\cap L_{p,U_{p}}$. In the same
way, we define the \textbf{kinematic radial velocity of }$\beta
^{\prime }$\textbf{\ with respect to }$\beta $, denoted by
$V_{\mathrm{kin}}^{\mathrm{rad}}$, and the \textbf{kinematic
tangential velocity of }$\beta ^{\prime }$\textbf{\ with respect
to }$\beta $, denoted by $V_{\mathrm{kin}}^{\mathrm{tng}}$.

We will say that $\beta $ is \textbf{kinematically comoving} with
$\beta ^{\prime }$ if $V_{\mathrm{kin}}=0$.
\end{definition}

Let $V_{\mathrm{kin}}^{\prime }$ be the kinematic relative
velocity of $\beta $ with respect to $\beta ^{\prime }$. Then,
$V_{\mathrm{kin}}=0$ if and only if $V_{\mathrm{kin}}^{\prime
}=0$, i.e. the relation \textquotedblleft to be kinematically
comoving with\textquotedblright\ is symmetric and so, we can say
that $\beta $ and $\beta ^{\prime }$ are kinematically comoving
(each one with respect to the other). Note that it is not
transitive in general.

\subsection{Fermi relative velocity}

We are going to define the ``Fermi relative velocity" as the
variation of the relative position.

\begin{definition}
\label{deffermi} Let $\beta $, $\beta ^{\prime }$ be two
observers, let $U$ be the 4-velocity of $\beta $, and let $S$ be
the relative position of $\beta ^{\prime }$ with respect to $\beta
$. The \textbf{Fermi relative velocity of }$\beta ^{\prime
}$\textbf{\ with respect to }$\beta $ is the projection of $\nabla
_{U}S$ onto $U^{\bot }$, i.e. it is the vector field
\begin{equation}
V_{\mathrm{Fermi}}:=\nabla _{U}S+g\left( \nabla _{U}S,U\right) U
\label{defvFermiusgusuu}
\end{equation}
defined on $\beta $. The right-hand side of
(\ref{defvFermiusgusuu}) is known as the \textit{Fermi
derivative}. The \textbf{Fermi radial velocity of }$\beta ^{\prime
}$\textbf{\ with respect to }$\beta $ is the component of
$V_{\mathrm{Fermi}}$ parallel to $S$, i.e.
$V_{\mathrm{Fermi}}^{\mathrm{rad}}:=g\left(
V_{\mathrm{Fermi}},\frac{S}{\left\Vert S\right\Vert }\right)
\frac{S}{\left\Vert S\right\Vert }$ if $S\neq 0$; if $S_{p}=0$
(i.e. $\beta $ and $\beta ^{\prime }$ intersect at $p$) then
$V_{\mathrm{Fermi}~p}^{\mathrm{rad}}:=V_{\mathrm{Fermi}~p}$. On
the other hand, the \textbf{Fermi tangential velocity of }$\beta
^{\prime }$\textbf{\ with respect to }$\beta $ is the component of
$V_{\mathrm{Fermi}}$ orthogonal to $S$, i.e.
$V_{\mathrm{Fermi}}^{\mathrm{tng}}:=V_{\mathrm{Fermi}}-V_{\mathrm{Fermi}}^{\mathrm{rad}}$.

We will say that $\beta $ is \textbf{Fermi-comoving} with $\beta
^{\prime }$ if $V_{\mathrm{Fermi}}=0$.
\end{definition}

It is important to remark that the modulus of the vectors of
$V_{\mathrm{Fermi}}$ is not necessarily smaller than one.

Since $g\left( V_{\mathrm{Fermi}},S\right) =g\left( \nabla
_{U}S,S\right) $, if $S\neq 0$ we have
\begin{equation}
V_{\mathrm{Fermi}}^{\mathrm{rad}}=g\left( \nabla
_{U}S,\frac{S}{\left\Vert S\right\Vert }\right)
\frac{S}{\left\Vert S\right\Vert }. \label{vradFermi}
\end{equation}
%So, the Fermi radial velocity of $\beta ^{\prime }$ with
%respect to $\beta $ has always full physical sense as the radial
%component of the variation of $S$ along the world line of the
%observer $\beta $, even if $\beta $ is not geodesic. This fact is
%also supported by Proposition \ref{propvradast2}, as we will see
%later.

The relation \textquotedblleft to be Fermi-comoving
with\textquotedblright\ is not symmetric in general.

An expression similar to (\ref{defvFermiusgusuu}) is given by the
next proposition, that can be proved easily.

\begin{proposition}
\label{propgusu}Let $\beta $, $\beta ^{\prime }$ be two observers,
let $U$ be the 4-velocity of $\beta $, let $S$ be the relative
position of $\beta ^{\prime }$ with respect to $\beta $, and let
$V_{\mathrm{Fermi}}$ be the Fermi relative velocity of $\beta
^{\prime }$ with respect to $\beta $. Then
$V_{\mathrm{Fermi}}=\nabla _{U}S-g\left( S,\nabla _{U}U\right) U$.
Note that if $\beta $ is geodesic, then $\nabla _{U}U=0$, and
hence $V_{\mathrm{Fermi}}=\nabla _{U}S$ .
\end{proposition}

If $S_{p}=0$, i.e. $\beta $ and $\beta ^{\prime }$ intersect at
$p$, then $V_{\mathrm{Fermi}~p}=\left( \nabla _{U}S\right) _{p}$.
So, it does not coincide in general with the concept of relative
velocity given in expression (\ref{relvel}).

We are going to introduce a concept of distance from the concept
of relative position given in Definition \ref{relpossp}. This
concept of distance was previously introduced by Fermi.

\begin{definition}
Let $u$ be an observer at an event $p$. Given $q$, $q^{\prime }\in
L_{p,u}$, and $s$, $s^{\prime }$ the relative positions of $q$,
$q^{\prime }$ with respect to $u$ respectively, the \textbf{Fermi
distance from }$q$\textbf{\ to }$q^{\prime }$\textbf{\ with
respect to }$u$ is the modulus of $s-s^{\prime }$, i.e.
$d_{u}^{\mathrm{Fermi}}\left( q,q^{\prime }\right) :=\left\Vert
s-s^{\prime }\right\Vert $.
\end{definition}

We have that $d_{u}^{\mathrm{Fermi}}$ is symmetric,
positive-definite and satisfies the triangular inequality. So, it
has all the properties that must verify a topological distance
defined on $L_{p,u}$. As a particular case, if $q^{\prime }=p$ we
have
\begin{equation}
d_{u}^{\mathrm{Fermi}}\left( q,p\right) =\left\Vert s\right\Vert
=\left( g\left( \exp _{p}^{-1}q,\exp _{p}^{-1}q\right) \right)
^{1/2}. \label{duspaceqp}
\end{equation}

The next proposition shows that the concept of Fermi distance is
the arc\-length parameter of a spacelike geodesic, and it can be
proved taking into account the properties of the exponential map
(see \cite{Helg62}).

\begin{proposition}
\label{propdspaceparam}Let $u$ be an observer at an event $p$.
Given $q\in L_{p,u}$ and $\alpha $ the unique geodesic from $p$ to
$q$, if we parameterize $\alpha $ by its arclength such that
$\alpha \left( 0\right) =p$, then $\alpha \left(
d_{u}^{\mathrm{Fermi}}\left( q,p\right) \right) =q$.
\end{proposition}

\begin{definition}
\label{defdistbeta2}Let $\beta $, $\beta ^{\prime }$ be two
observers and let $S$ be the relative position of $\beta ^{\prime
}$\ with respect to $\beta $. The \textbf{Fermi distance from
}$\beta ^{\prime }$\textbf{\ to }$\beta $\textbf{\ with respect to
}$\beta $ is the scalar field $\left\Vert S\right\Vert $ defined
in $\beta $.
\end{definition}

We are going to characterize the Fermi radial velocity in terms of
the Fermi distance.

\begin{proposition}
\label{propvradast2}Let $\beta $, $\beta ^{\prime }$ be two
observers, let $S$ be the relative position of $\beta ^{\prime }$
with respect to $\beta $, and let $U$ be the 4-velocity of $\beta
$. If $S\neq 0$, the Fermi radial velocity of $\beta ^{\prime }$\
with respect to $\beta $ reads
$V_{\mathrm{Fermi}}^{\mathrm{rad}}=U\left( \left\Vert S\right\Vert
\right) \frac{S}{\left\Vert S\right\Vert }$.
\end{proposition}

By Definition \ref{defdistbeta2} and Proposition
\ref{propvradast2}, the Fermi radial velocity of $\beta ^{\prime
}$ with respect to $\beta $ is the rate of change of the Fermi
distance from $\beta ^{\prime }$ to $\beta $ with respect to
$\beta $. So, if we parameterize $\beta $ by its proper time $\tau
$, the Fermi radial velocity of $\beta ^{\prime }$ with respect to
$\beta $ at $p=\beta \left( \tau _{0}\right) $ is given by
$V_{\mathrm{Fermi}~p}^{\mathrm{rad}}=\frac{\mathrm{d}\left(
\left\Vert S\right\Vert \circ \beta \right) }{\mathrm{d}\tau
}\left( \tau _{0}\right) \frac{S_{p}}{\left\Vert S_{p}\right\Vert
}$, where $\left\Vert S\right\Vert \circ \beta$ is the Fermi distance as a function of $\tau$.

\section{Relative velocity in the framework of lightlike simultaneity}
\label{sec4}

The lightlike (or observed) simultaneity is based on ``what an
observer is really observing" and it provides an appropriate
framework to study optical phenomena and observational cosmology
(see \cite{Elli85}).

Let $p\in \mathcal{M}$ and $\varphi :\mathcal{M}\rightarrow
\mathbb{R}$ defined by $\varphi \left( q\right) :=g\left( \exp
_{p}^{-1}q,\exp _{p}^{-1}q\right) $. Then, it is a submersion and
the set
\begin{equation}
E_{p}:=\varphi ^{-1}\left( 0\right) -\left\{ p\right\}
\label{f1.horismos}
\end{equation}%
is a regular 3-dimensional submanifold, called \textit{horismos
submanifold of }$p$ (see \cite{Bolo02}, \cite{Beem81}). An event
$q$ is in $E_{p}$ if and only if $q\neq p$ and there exists a
lightlike geodesic joining $p$ and $q$. $E_{p}$ has two connected
components, $E_{p}^{-}$ and $E_{p}^{+}$ \cite{Sach77}; $E_{p}^{-}$
(respectively $E_{p}^{+}$) is the \textit{past-pointing}
(respectively \textit{future-pointing}) \textit{horismos
submanifold of }$p$, and it is the connected component of
(\ref{f1.horismos}) in which, for each event $q\in E_{p}^{-}$
(respectively $q\in E_{p}^{+}$), the preimage $\exp _{p}^{-1}q$ is
a past-pointing (respectively future-pointing) lightlike vector.
In other words, $E_{p}^{-}=\exp _{p}C_{p}^{-}$, and
$E_{p}^{+}=\exp _{p}C_{p}^{+}$, where $C_{p}^{-}$ and $C_{p}^{+}$
are the past-pointing and the future-pointing light cones of
$T_{p}\mathcal{M}$ respectively.

This section is analogous to Section \ref{sec3}, but using
$E_{p}^{-}$ instead of $L_{p,u}$.

\begin{figure}[tbp]
\begin{center}
\includegraphics[width=0.7\textwidth]{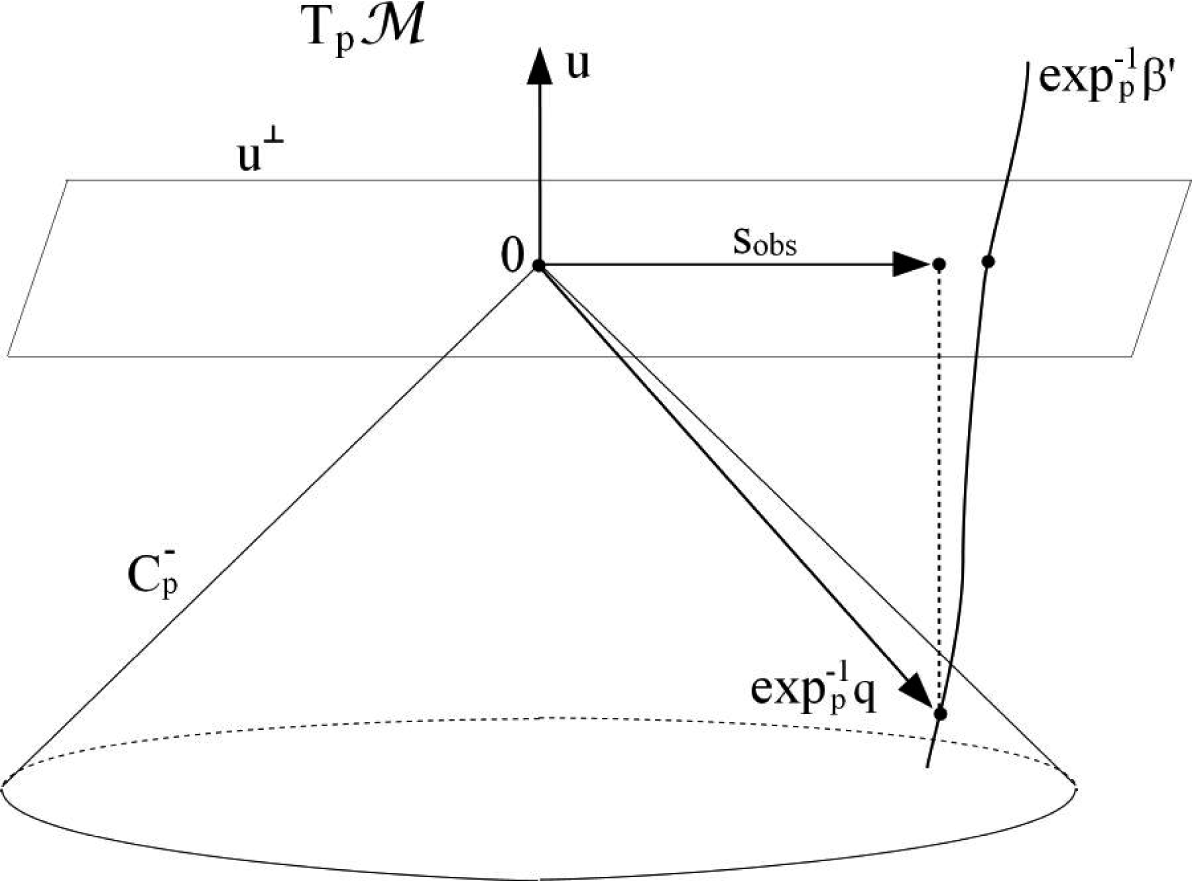}
\end{center}
\caption{Scheme in $T_p \mathcal{M}$ of the relative position
$s_{\mathrm{obs}}$ of $q$ observed by $u$.} \label{figrelposobs}
\end{figure}

\begin{definition}
\label{defrq}Given $u$ an observer at $p$, and an observed event
$q\in E_{p}^{-}\cup \left\{ p\right\} $, the \textbf{relative
position of }$q$\textbf{\ observed by }$u$ (or the observed
relative position of $q$ with respect to $u$) is the projection of
$\exp _{p}^{-1}q$ onto $u^{\bot }$ (see Fig. \ref{figrelposobs}),
i.e. $s_{\mathrm{obs}}:=\exp _{p}^{-1}q+g\left( \exp
_{p}^{-1}q,u\right) u$.
\end{definition}

We can generalize this definition for two observers $\beta $ and
$\beta ^{\prime }$.

\begin{definition}
\label{defsobs} Let $\beta $, $\beta ^{\prime }$ be two observers
and let $U$ be the 4-velocity of $\beta $. The \textbf{relative
position of }$\beta ^{\prime }$\textbf{\ observed by }$\beta $ is
the vector field $S_{\mathrm{obs}}$ defined in $\beta $ such that
$S_{\mathrm{obs}~p}$ is the relative position of $q$ observed by
$U_{p}$, where $p\in \beta $ and $q$ is the unique event of $\beta
^{\prime }\cap E_{p}^{-}$.
\end{definition}

\subsection{Spectroscopic relative velocity}

In a previous work (see \cite{Bolo05}), we defined a concept of
relative velocity of an observer observed by another observer in
the framework of lightlike simultaneity. We are going to rename
this concept as \textquotedblleft spectroscopic relative
velocity\textquotedblright , and to review its properties in the
context of this work.

\begin{figure}[tbp]
\begin{center}
\includegraphics[width=0.3\textwidth]{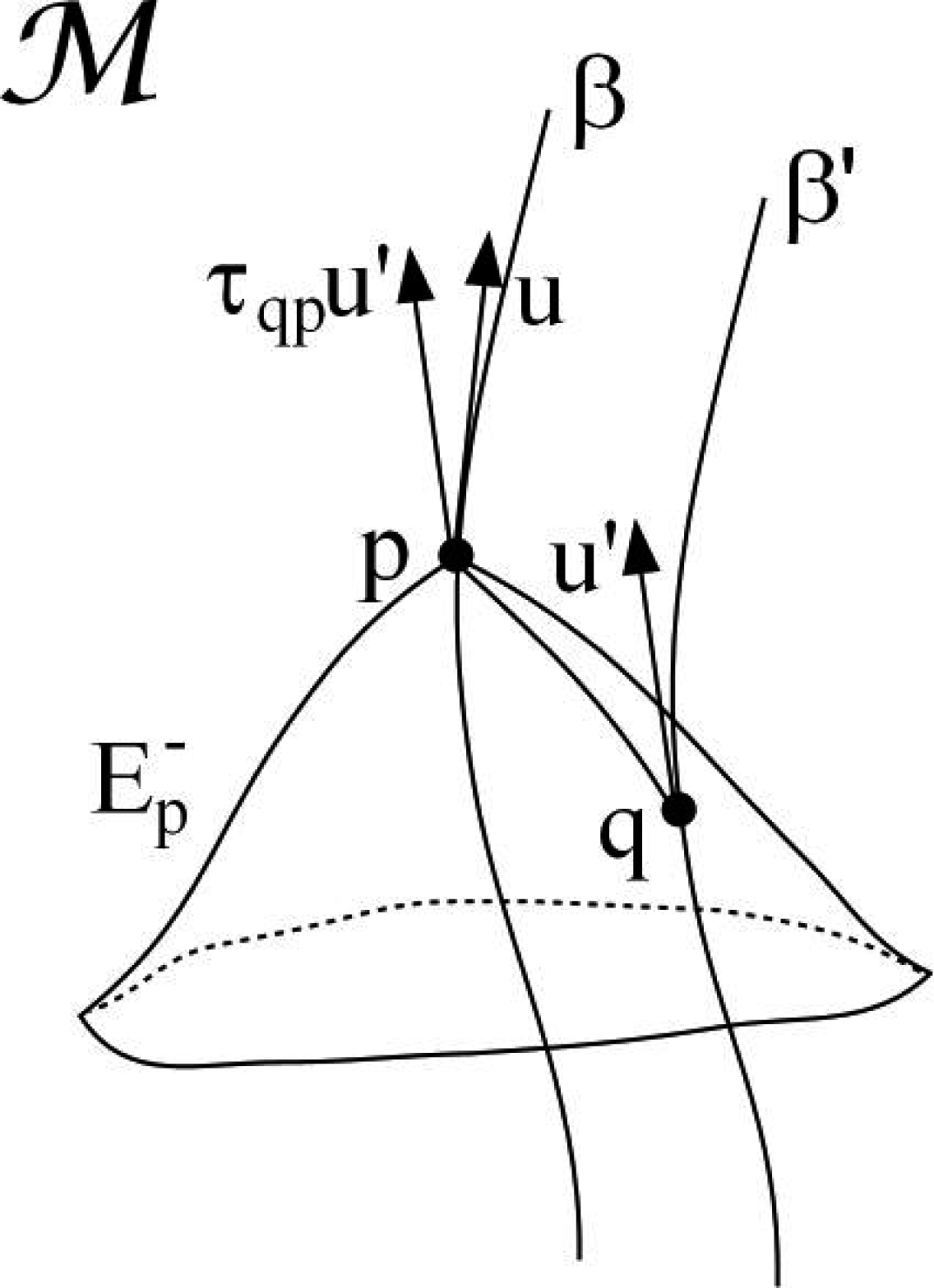}
\end{center}
\caption{Scheme in $\mathcal{M}$ of the elements that involve the
definition of the spectroscopic relative velocity of $u^{\prime}$
observed by $u$.} \label{figvspec}
\end{figure}

\begin{definition}
\label{relvelu}Let $u$, $u^{\prime }$ be two observers at $p$, $q$
respectively such that $q\in E_{p}^{-}$ and let $\lambda $ be a
light ray from $q$ to $p$. The \textbf{spectroscopic relative
velocity of }$u^{\prime } $\textbf{\ observed by }$u$ is the
unique vector $v_{\mathrm{spec}}\in u^{\bot }$ such that $\tau
_{qp}u^{\prime }=\gamma \left( u+v_{\mathrm{spec}}\right) $, where
$\gamma $ is the gamma factor corresponding to the velocity $\Vert
v_{\mathrm{spec}}\Vert $ (see Fig. \ref{figvspec}). So, it is
given by
\begin{equation}
v_{\mathrm{spec}}:=\frac{1}{-g\left( \tau _{qp}u^{\prime
},u\right) }\tau _{qp}u^{\prime }-u.  \label{vspec}
\end{equation}

We define the \textbf{spectroscopic radial} and \textbf{tangential
velocity of }$u^{\prime }$\textbf{\ observed by }$u$ analogously
to Definition \ref{kinrelvel}, using $s_{\mathrm{obs}}$ (see
Definition \ref{defrq}) instead of $s$.

%Let $s_{\mathrm{obs}}$ be the relative position of $q$ observed by
%$u$, the \textbf{spectroscopic radial velocity of }$u^{\prime
%}$\textbf{\ observed by }$u$ is the component of
%$v_{\mathrm{spec}}$ parallel to $s_{\mathrm{obs}}$, i.e.
%$v_{\mathrm{spec}}^{\mathrm{rad}}:=g\left(
%v_{\mathrm{spec}},\frac{s_{\mathrm{obs}}}{\left\Vert
%s_{\mathrm{obs}}\right\Vert }\right)
%\frac{s_{\mathrm{obs}}}{\left\Vert s_{\mathrm{obs}}\right\Vert }$.
%If $s_{\mathrm{obs}}=0$ (i.e. $p=q$) then
%$v_{\mathrm{spec}}^{\mathrm{rad}}:=v_{\mathrm{spec}}$. On the
%other hand, the \textbf{spectroscopic tangential velocity of
%}$u^{\prime }$\textbf{\ observed by }$u$ is the component of
%$v_{\mathrm{spec}}$ orthogonal to $s_{\mathrm{obs}}$, i.e.
%$v_{\mathrm{spec}}^{\mathrm{tng}}:=v_{\mathrm{spec}}-v_{\mathrm{spec}}^{\mathrm{rad}}$.
\end{definition}

So, the spectroscopic relative velocity of $u^{\prime }$ observed
by $u$ is the relative velocity of $\tau _{qp}u^{\prime }$
observed by $u$, in the sense of expression (\ref{relvel}), and
$\Vert v_{\mathrm{spec}}\Vert <1$.

Note that if $w$ is the relative velocity of $\lambda $ observed
by $u$ (see (\ref{f1.efepe})), then
$w=-\frac{s_{\mathrm{obs}}}{\left\Vert s_{\mathrm{obs}}\right\Vert
}$, and so
\begin{equation}
v_{\mathrm{spec}}^{\mathrm{rad}}=g\left(
v_{\mathrm{spec}},w\right) w. \label{vradspecw}
\end{equation}

We can generalize these definitions for two observers $\beta $ and
$\beta ^{\prime }$.

\begin{definition}
\label{defspecvel} Let $\beta $, $\beta ^{\prime }$ be two
observers, we define $V_{\mathrm{spec}}$ (the
\textbf{spectroscopic relative velocity of }$\beta ^{\prime
}$\textbf{\ observed by }$\beta $) and its \textbf{radial} and
\textbf{tangential} components analogously to Definition
\ref{defkinrelvel}, using $E_{p}^{-}$ instead of $L_{p,U_p}$.

We will say that $\beta $ is \textbf{spectroscopically comoving}
with $\beta ^{\prime }$ if $V_{\mathrm{spec}}=0$.

%Let $\beta $, $\beta ^{\prime }$ be two observers, and let $U$,
%$U^{\prime }$ be the $4$-velocities of $\beta $, $\beta ^{\prime
%}$ respectively. The \textbf{spectroscopic relative velocity of
%}$\beta ^{\prime }$\textbf{\ observed by }$\beta $ is the vector
%field $V_{\mathrm{spec}}$ defined on $\beta $ such that
%$V_{\mathrm{spec}~p}$ is the spectroscopic relative velocity of
%$U_{q}^{\prime }$ observed by $U_{p}$ (in the sense of Definition
%\ref{relvelu}), where $p\in \beta $ and $q$ is the unique event of
%$\beta ^{\prime }\cap E_{p}^{-}$. In the same way, we define the
%\textbf{spectroscopic radial velocity of }$\beta ^{\prime
%}$\textbf{\ observed by }$\beta $, denoted by
%$V_{\mathrm{spec}}^{\mathrm{rad}}$, and the \textbf{spectroscopic
%tangential velocity of }$\beta ^{\prime }$\textbf{\ observed by
%}$\beta $, denoted by $V_{\mathrm{spec}}^{\mathrm{tng}}$. We will
%say that $\beta $ is \textbf{spectroscopically comoving} with
%$\beta ^{\prime }$ if $V_{\mathrm{spec}}=0$.
\end{definition}

The relation \textquotedblleft to be spectroscopically comoving
with\textquotedblright\ is not symmetric in general.

The following result can be found in \cite{Bolo05}.

\begin{proposition}
Let $\lambda $ be a light ray from $q$\ to $p$ and let $u$,
$u^{\prime }$\ be two observers at $p$, $q$\ respectively. Then
\begin{equation}
\nu ^{\prime }=\gamma \left( 1-g\left( v_{\mathrm{spec}},w\right)
\right) \nu , \label{dopplergen}
\end{equation}
where $\nu $, $\nu ^{\prime }$ are the frequencies of $\lambda $
observed by $u$, $u^{\prime }$ respectively, $v_{\mathrm{spec}}$
is the spectroscopic relative velocity of $u^{\prime }$ observed
by $u$, $w$ is the relative velocity of $\lambda $ observed by
$u$, and $\gamma $ is the gamma factor corresponding to the
velocity $\Vert v_{\mathrm{spec}}\Vert $.
\end{proposition}

Expression (\ref{dopplergen}) is the general expression for
Doppler effect (that includes gravitational redshift, see
\cite{Bolo05}). Therefore, if $\beta $ is spectroscopically
comoving with $\beta ^{\prime }$, and $\lambda $ is a light ray
from $\beta ^{\prime }$ to $\beta $, then, by (\ref{dopplergen}),
we have that $\beta $ and $\beta ^{\prime }$ observe $\lambda $
with the same frequency. So, if $\beta '$ emits $n$ light rays in
a unit of its proper time, then $\beta $ observes also $n$ light
rays in a unit of its proper time. Hence, $\beta $ observes that
$\beta '$ uses the \textquotedblleft same clock\textquotedblright\
as its.

Taking into account (\ref{vradspecw}), expression
(\ref{dopplergen}) can be written in the form
\begin{equation}
\label{fshiftvrad} \nu ^{\prime }=\frac{1\pm \Vert
v_{\mathrm{spec}}^{\mathrm{rad}}\Vert }{\sqrt{1-\Vert
v_{\mathrm{spec}}\Vert ^{2}}}\nu ,
\end{equation}
where we choose ``$+$" if $g\left( v_{\mathrm{spec}},w\right) <0$
(i.e. if $u^{\prime }$ is moving away from $u$), and we choose
``$-$" if $g\left( v_{\mathrm{spec}},w\right) >0$ (i.e. if
$u^{\prime }$ is getting closer to $u$).

\begin{remark}
\label{remark1} We can not deduce $v_{\mathrm{spec}}$ from the
shift, $\nu ^{\prime }/\nu $, unless we make some assumptions
(like considering negligible the tangential component of
$v_{\mathrm{spec}}$, as we will see in Remark \ref{rem2}). For
instance, if $\nu ^{\prime }/\nu =1$ then $v_{\mathrm{spec}}$ is
not necessarily zero. Let us study this particular case: by
(\ref{dopplergen}) we have
\[
1=\frac{\nu ^{\prime }}{\nu }=\frac{1-g\left(
v_{\mathrm{spec}},w\right) }{\sqrt{1-\Vert v_{\mathrm{spec}}\Vert
^{2}}}\longrightarrow g\left( v_{\mathrm{spec}},w\right)
=1-\sqrt{1-\Vert v_{\mathrm{spec}}\Vert ^{2}}.
\]
Since $\left( 1-\sqrt{1-\Vert v_{\mathrm{spec}}\Vert ^{2}}\right)
\geq 0$, it is necessary that $g\left( v_{\mathrm{spec}},w\right)
\geq 0$, i.e. the observed object has to be getting closer to the
observer. In this case, by (\ref{fshiftvrad}) we have $\Vert
v_{\mathrm{spec}}^{\mathrm{rad}}\Vert =1-\sqrt{1-\Vert
v_{\mathrm{spec}}\Vert ^{2}}$. So, it is possible that $\nu '/\nu
=1$ and $v_{\mathrm{spec}}\neq 0$ if the observed object is
getting closer to the observer. On the other hand, if the observed
object is moving away from the observer then $\nu ^{\prime }/\nu
=1$ if and only if $v_{\mathrm{spec}}=0$. That is, for objects
moving away, the shift is always redshift; and for objects getting
closer, the shift can be blueshift, 1, or redshift.
\end{remark}

\begin{remark}
\label{rem2}If we suppose that
$v_{\mathrm{spec}}^{\mathrm{tng}}=0$, i.e.
$v_{\mathrm{spec}}=v_{\mathrm{spec}}^{\mathrm{rad}}=kw$ with $k\in
\left] -1,1\right[ $, then we can deduce $v_{\mathrm{spec}}$ from
the shift $\nu ^{\prime }/\nu $:
\[
\frac{\nu ^{\prime }}{\nu }=\frac{1-g\left(
v_{\mathrm{spec}},w\right) }{\sqrt{1-\Vert v_{\mathrm{spec}}\Vert
^{2}}}=\frac{1-k}{\sqrt{1-k^{2}}}=\frac{\sqrt{1-k}}{\sqrt{1+k}}\longrightarrow
k=\frac{1-\left( \frac{\nu ^{\prime }}{\nu }\right) ^{2}}{1+\left(
\frac{\nu ^{\prime }}{\nu }\right) ^{2}},
\]
and hence
\begin{equation}
v_{\mathrm{spec}}=\left( \frac{1-\left( \frac{\nu ^{\prime }}{\nu
}\right) ^{2}}{1+\left( \frac{\nu ^{\prime }}{\nu }\right)
^{2}}\right) w=-\left( \frac{1-\left( \frac{\nu ^{\prime }}{\nu
}\right) ^{2}}{1+\left( \frac{\nu ^{\prime }}{\nu }\right)
^{2}}\right) \frac{s_{\mathrm{obs}}}{\left\Vert
s_{\mathrm{obs}}\right\Vert }. \label{vspecshift}
\end{equation}
\end{remark}

\subsection{Astrometric relative velocity}

We are going to define the ``astrometric relative velocity" as the
variation of the observed relative position.

\begin{definition}
Let $\beta $, $\beta ^{\prime }$ be two observers, we define
$V_{\mathrm{ast}}$ (the \textbf{astrometric relative velocity of
}$\beta ^{\prime }$\textbf{\ observed by }$\beta $) and its
\textbf{radial} and \textbf{tangential} components analogously to
Definition \ref{deffermi}, using $S_{\mathrm{obs}}$ (see
Definition \ref{defsobs}) instead of $S$. So,
\begin{equation}
V_{\mathrm{ast}}:=\nabla _{U}S_{\mathrm{obs}}+g\left( \nabla
_{U}S_{\mathrm{obs}},U\right) U, \label{defvastusgusuu}
\end{equation}
where $U$ is the 4-velocity of $\beta $.

We will say that $\beta $ is \textbf{astrometrically comoving}
with $\beta ^{\prime }$ if $V_{\mathrm{ast}}=0$.

%Let $\beta $, $\beta ^{\prime }$ be two observers, let $U$ be the
%4-velocity of $\beta $, and let $S_{\mathrm{obs}}$ be the relative
%position of $\beta ^{\prime }$ observed by $\beta $. The
%\textbf{astrometric relative velocity of }$\beta ^{\prime
%}$\textbf{\ observed by }$\beta $ is the projection of $\nabla
%_{U}S_{\mathrm{obs}}$ onto $U^{\bot }$, i.e. it is the vector
%field
%\begin{equation}
%V_{\mathrm{ast}}:=\nabla _{U}S_{\mathrm{obs}}+g\left( \nabla
%_{U}S_{\mathrm{obs}},U\right) U \label{defvastusgusuu}
%\end{equation}
%defined on $\beta $. The \textbf{astrometric radial velocity of
%}$\beta ^{\prime }$\textbf{\ observed by }$\beta $ is the
%component of $V_{\mathrm{ast}}$ parallel to $S_{\mathrm{obs}}$,
%i.e. $V_{\mathrm{ast}}^{\mathrm{rad}}:=g\left(
%V_{\mathrm{ast}},\frac{S_{\mathrm{obs}}}{\left\Vert
%S_{\mathrm{obs}}\right\Vert }\right)
%\frac{S_{\mathrm{obs}}}{\left\Vert S_{\mathrm{obs}}\right\Vert }$
%if $S_{\mathrm{obs}}\neq 0$; if $S_{\mathrm{obs}~p}=0$ (i.e.
%$\beta $ and $\beta ^{\prime }$ intersect at $p$) then
%$V_{\mathrm{ast}~p}^{\mathrm{rad}}:=V_{\mathrm{ast}~p}$. On the
%other hand, the \textbf{astrometric tangential velocity of }$\beta
%^{\prime }$\textbf{\ observed by }$\beta $ is the component of
%$V_{\mathrm{ast}}$ orthogonal to $S_{\mathrm{obs}}$, i.e.
%$V_{\mathrm{ast}}^{\mathrm{tng}}:=V_{\mathrm{ast}}-V_{\mathrm{ast}}^{\mathrm{rad}}$.
%We will say that $\beta $ is \textbf{astrometrically comoving}
%with $\beta ^{\prime }$ if $V_{\mathrm{ast}}=0$.
\end{definition}

It is important to remark that the modulus of the vectors of
$V_{\mathrm{ast}}$ is not necessarily smaller than one.

Analogously to (\ref{vradFermi}), since $g\left(
V_{\mathrm{ast}},S_{\mathrm{obs}}\right) =g\left( \nabla
_{U}S_{\mathrm{obs}},S_{\mathrm{obs}}\right) $, if
$S_{\mathrm{obs}}\neq 0$ we have
\begin{equation}
V_{\mathrm{ast}}^{\mathrm{rad}}=g\left( \nabla
_{U}S_{\mathrm{obs}},\frac{S_{\mathrm{obs}}}{\left\Vert
S_{\mathrm{obs}}\right\Vert }\right)
\frac{S_{\mathrm{obs}}}{\left\Vert S_{\mathrm{obs}}\right\Vert }.
\label{vradast}
\end{equation}
%So, the astrometric radial velocity of $\beta ^{\prime }$ observed
%by $\beta $ has always full physical sense as the radial component
%of the variation of $S_{\mathrm{obs}}$ along the world line of the observer
%$\beta $, even if $\beta $ is not geodesic. This fact is also
%supported by Proposition \ref{propvradast}, as we will see later.

The relation \textquotedblleft to be astrometrically comoving
with\textquotedblright\ is not symmetric in general.

An expression similar to (\ref{defvastusgusuu}) is given by the
next proposition, which proof is analogous to the proof of
Proposition \ref{propgusu}.

\begin{proposition}
\label{prop10}
Let $\beta $, $\beta ^{\prime }$ be two observers, let $U$ be the
4-velocity of $\beta $, let $S_{\mathrm{obs}}$ be the relative
position of $\beta ^{\prime }$ observed by $\beta $, and let
$V_{\mathrm{ast}}$ be the astrometric relative velocity of $\beta
^{\prime }$ observed by $\beta $. Then $V_{\mathrm{ast}}=\nabla
_{U}S_{\mathrm{obs}}-g\left( S_{\mathrm{obs}},\nabla _{U}U\right)
U$. Note that if $\beta $ is geodesic, then $\nabla _{U}U=0$, and
hence $V_{\mathrm{ast}}=\nabla _{U}S_{\mathrm{obs}}$.
\end{proposition}

If $S_{\mathrm{obs}~p}=0$, i.e. $\beta $ and $\beta ^{\prime }$
intersect at $p$, then $V_{\mathrm{ast}~p}=\left( \nabla
_{U}S_{\mathrm{obs}}\right) _{p}$. So, it does not coincide in
general with the concept of relative velocity given in
(\ref{relvel}).

We are going to introduce another concept of distance from the
concept of observed relative position given in Definition
\ref{defrq}. This distance was previously introduced in
\cite{Kerm32} and studied in \cite{Bolo05}, and it plays a basic
role for the construction of \textit{optical coordinates} whose
relevance for cosmology was stressed in many articles by G. Ellis
and his school (see \cite{Elli85}).

\begin{definition}
Let $u$ be an observer at an event $p$. Given $q$, $q^{\prime }\in
E_{p}^{-}\cup \left\{ p\right\} $, and $s_{\mathrm{obs}}$,
$s_{\mathrm{obs}}^{\prime }$ the relative positions of $q$,
$q^{\prime }$ observed by $u$ respectively, the \textbf{affine
distance from }$q$\textbf{\ to }$q^{\prime }$\textbf{\ observed by
}$u$ is the modulus of $s_{\mathrm{obs}}-s_{\mathrm{obs}}^{\prime
}$, i.e. $d_{u}^{\mathrm{aff}}\left( q,q^{\prime }\right)
:=\left\Vert s_{\mathrm{obs}}-s_{\mathrm{obs}}^{\prime
}\right\Vert $.
\end{definition}

We have that $d_{u}^{\mathrm{aff}}$ is symmetric,
positive-definite and satisfies the triangular inequality. So, it
has all the properties that must verify a topological distance
defined on $E_{p}^{-}\cup \left\{ p\right\} $. As a particular
case, if $q^{\prime }=p$ we have
\begin{equation}
d_{u}^{\mathrm{aff}}\left( q,p\right) =\left\Vert
s_{\mathrm{obs}}\right\Vert =g\left( \exp _{p}^{-1}q,u\right) .
\label{dulightqp}
\end{equation}

The next proposition shows that the concept of affine distance is
according to the concept of ``length" (or ``time") parameter of a
lightlike geodesic for an observer, and it is proved in
\cite{Bolo05}.

\begin{proposition}
\label{propdlightparam}Let $\lambda $ be a light ray from $q$ to
$p$, let $u$ be an observer at $p$, and let $w$ be the relative
velocity of $\lambda $ observed by $u$. If we parameterize
$\lambda $ affinely (i.e. the vector field tangent to $\lambda $
is parallelly transported along $\lambda $) such that $\lambda
\left( 0\right) =p$ and $\overset{.}{\lambda }\left( 0\right)
=-\left( u+w\right) $, then $\lambda \left(
d_{u}^{\mathrm{aff}}\left( q,p\right) \right) =q$.
\end{proposition}

\begin{definition}
\label{defdistbeta}Let $\beta $, $\beta ^{\prime }$ be two
observers and let $S_{\mathrm{obs}}$ be the relative position of
$\beta ^{\prime }$\ observed by $\beta $. The \textbf{affine
distance from }$\beta ^{\prime }$\textbf{\ to }$\beta $\textbf{\
observed by }$\beta $ is the scalar field $\left\Vert
S_{\mathrm{obs}}\right\Vert $ defined in $\beta $.
\end{definition}

We are going to characterize the astrometric radial velocity in
terms of the affine distance. The proof of the next proposition is
analogous to the proof of Proposition \ref{propvradast2}, taking
into account expression (\ref{vradast}).

\begin{proposition}
\label{propvradast}Let $\beta $, $\beta ^{\prime }$ be two
observers, let $S_{\mathrm{obs}}$ be the relative position of
$\beta ^{\prime }$ observed by $\beta $, and let $U$ be the
4-velocity of $\beta $. If $S_{\mathrm{obs}}\neq 0$, the
astrometric radial velocity of $\beta ^{\prime }$\ observed by
$\beta $ reads $V_{\mathrm{ast}}^{\mathrm{rad}}=U\left( \left\Vert
S_{\mathrm{obs}}\right\Vert \right)
\frac{S_{\mathrm{obs}}}{\left\Vert S_{\mathrm{obs}}\right\Vert }$.
\end{proposition}

By Definition \ref{defdistbeta} and Proposition \ref{propvradast},
the astrometric radial velocity of $\beta ^{\prime }$ observed by
$\beta $ is the rate of change of the affine distance from $\beta
^{\prime }$ to $\beta $ observed by $\beta $. So, if we
parameterize $\beta $ by its proper time $\tau $, the astrometric
radial velocity of $\beta ^{\prime }$ observed by $\beta $ at
$p=\beta \left( \tau _{0}\right) $ is given by
$V_{\mathrm{ast}~p}^{\mathrm{rad}}=\frac{\mathrm{d}\left(
\left\Vert S_{\mathrm{obs}}\right\Vert \circ \beta \right)
}{\mathrm{d}\tau }\left( \tau _{0}\right)
\frac{S_{\mathrm{obs}~p}}{\left\Vert S_{\mathrm{obs}~p}\right\Vert
}$, where $\left\Vert S_{\mathrm{obs}}\right\Vert \circ \beta$ is the affine distance as a function of $\tau$.

\section{Special Relativity}
\label{sec5}

In this section, we are going to work in the Minkowski spacetime,
considering $\beta $, $\beta '$ two observers with 4-velocities $U$, $U'$ respectively. The goal is to find expressions for $V_{\mathrm{Fermi}}$ and $V_{\mathrm{ast}}$ in terms of
$U$, $\nabla _U U$, $U'$, $S$ and $S_{\mathrm{obs}}$, i.e. without $\nabla _U S$, $\nabla _U S_{\mathrm{obs}}$, or any term involving the evolution of $S$, $S_{\mathrm{obs}}$.

\begin{proposition}
\label{propmink1} Let $S$ be the relative position of $\beta
'$ with respect to $\beta $, and let
$V_{\mathrm{Fermi}}$ be the Fermi relative velocity
of $\beta '$ with respect to $\beta $. Then
\begin{equation}
\label{fmink1_3}
V_{\mathrm{Fermi}} =\left( 1+g\left( S,\nabla _{U}U\right)
\right) \left( \frac{1}{-g\left( U',U\right)
}U'-U\right) ,
\end{equation}
where $V_{\mathrm{Fermi}}$, $U$, $S$, $\nabla _U U$ are evaluated at an event $p$ of $\beta $, and $U'$ is evaluated at the event $q=\beta '\cap L_{p,U_p}$.
\end{proposition}

\begin{proof}
We are going to consider the observers parameterized by their
proper times. Let $p=\beta \left( \tau \right) $ be an event of
$\beta $, let $u\left( \tau \right) $ be the 4-velocity of $\beta
$ at $p$, and let $q=\beta '\left( \tau '\left( \tau \right) \right)
$ be the event of $\beta '$ such that $g\left( u\left(
\tau \right) ,q-p\right) =0$ (note that the Minkowski spacetime
has an affine structure, and $q-p$ denotes the vector which joins
$p$ and $q$). So, $\tau '\left( \tau \right) $ is the proper time of $q=\beta '\cap L_{p,u}$, and the relative position of $q$ with
respect to $u$, denoted by $s$, is $q-p$. If $u'\left( \tau '\right) $ is the 4-velocity of $\beta '$ at $q$, then
\begin{equation}
\label{eqm1}
s\left( \tau \right) =\beta' \left( \tau '\left( \tau \right) \right)-\beta \left( \tau \right) \,\Longrightarrow \, \dot{s}=u'\left( \tau '\right) \dot{\tau }'-u,
\end{equation}
where the dot denotes $\frac{d}{d\tau }$.
On the other hand
\begin{equation}
\label{eqm2}
g\left( s,u\right) =0 \,\Longrightarrow \, g\left( \dot{s},u\right) +g\left( s,\dot{u}\right) =0.
\end{equation}
Applying (\ref{eqm1}) in (\ref{eqm2}) we have
\begin{equation}
\label{eqm3}
g\left( u'\left( \tau '\right) \dot{\tau }'-u,u\right) +g\left( s,\dot{u}\right) =0 \,\Longrightarrow \, \dot{\tau }'=\frac{1+g\left( s,\dot{u}\right) }{-g\left( u'\left( \tau '\right) ,u\right) }.
\end{equation}
Combining (\ref{eqm1}) and (\ref{eqm3}), we obtain
\begin{equation}
\label{fmink1}
\dot{s}=\frac{1+g\left( s,\dot{u}\right) }{-g\left( u'\left( \tau '\right) ,u\right) }u'\left( \tau '\right) -u.
\end{equation}

Let $U$, $U'$ be the 4-velocities of $\beta $ and $\beta
'$ respectively, and let $S$ be the relative position of
$\beta '$ with respect to $\beta $. Then, from
(\ref{fmink1}) we have
\begin{equation}
\label{fmink1_2}
\nabla _{U}S=\frac{1+g\left( S,\nabla _{U}U\right) }{-g\left( U',U\right) }U'-U,
\end{equation}
where $U$, $S$, $\nabla _U U$, $\nabla _U S$ are evaluated at $p$, and $U'$ is evaluated at $q$.
So, by Proposition \ref{propgusu} and expression (\ref{fmink1_2}),
the Fermi relative velocity $V_{\mathrm{Fermi}}$ of $\beta
'$ with respect to $\beta $ is given by
\begin{eqnarray*}
V_{\mathrm{Fermi}} &=&\nabla _U S-g\left(
S,\nabla _U U\right) U  \nonumber \\
&=&\left( 1+g\left( S,\nabla _U U\right)
\right) \left( \frac{1}{-g\left( U',U\right)
}U'-U\right) ,
\end{eqnarray*}
where $V_{\mathrm{Fermi}}$, $U$, $S$, $\nabla _U U$ are evaluated at $p$, and $U'$ is evaluated at $q$.
\end{proof}

Taking into account the expression of the kinematic relative velocity given in (\ref{fkinrelvel}), we obtain the next corollary:

\begin{corollary}
The Fermi relative velocity of $\beta '$ with respect to $\beta $ reads
\begin{equation}
\label{fmink1_3rel}
V_{\mathrm{Fermi}} =\left( 1+g\left( S,\nabla _U U\right) \right) V_{\mathrm{kin}}.
\end{equation}
\end{corollary}

So, $V_{\mathrm{Fermi}}$ and $V_{\mathrm{kin}}$ are proportional.
Moreover, if $\beta $ is geodesic, then
$V_{\mathrm{Fermi}}=V_{\mathrm{kin}}$.

\begin{proposition}
\label{propmink2}
Let $S_{\mathrm{obs}}$ be the relative position of $\beta
'$ observed by $\beta $, and let
$V_{\mathrm{ast}}$ be the astrometric relative velocity
of $\beta '$ with respect to $\beta $. Then
\begin{equation}
\label{fmink2_3}
V_{\mathrm{ast}} = \frac{1}{g\left( U',\frac{S_{\mathrm{obs}}}{\| S_{\mathrm{obs}}\| }-U\right) }\left( U'+g\left( U',U\right) U\right) +\| S_{\mathrm{obs}}\| \nabla _U U,
\end{equation}
where $V_{\mathrm{ast}}$, $U$, $S_{\mathrm{obs}}$, $\nabla _U U$ are evaluated at an event $p$ of $\beta $, and $U'$ is evaluated at the event $q=\beta '\cap E^-_p$.
\end{proposition}

\begin{proof}

We are going to consider the observers parameterized by their
proper times. Let $p=\beta \left( \tau \right) $ be an event of
$\beta $, let $u\left( \tau \right) $ be the 4-velocity of $\beta
$ at $p$, and let $q=\beta '\left( \tau '\left( \tau \right) \right)
$ be the event of $\beta '$ such that $g\left( q-p,q-p\right) =0$ (note that the Minkowski spacetime
has an affine structure, and $q-p$ denotes the vector which joins
$p$ and $q$). So, $\tau '\left( \tau \right) $ is the proper time of $q=\beta ' \cap E^-_p$, and the
relative position of $q$ observed by $u$, denoted by $s_{\textrm{obs}}$, is the projection of $q-p$ onto $u^{\bot }$. Let us denote $s_{\mathrm{obs}}$ by $s$ for the shake of readability. Hence
\begin{equation}
\label{eqm21}
s\left( \tau \right) =\beta' \left( \tau '\left( \tau \right) \right)-\beta \left( \tau \right) +\| s\left( \tau\right) \| u,
\end{equation}
where $\| s\| =\sqrt{g\left( s,s\right) }$ is the affine distance from $p$ to $q$.
If $u'\left( \tau '\right) $ is the 4-velocity of $\beta '$ at $q$, deriving (\ref{eqm21}) with respect to $\tau$ we obtain
\begin{equation}
\label{eqm22}
\dot{s}=u'\left( \tau '\right) \dot{\tau }'-u+g\left( \dot{s},\frac{s}{\| s\| }\right) u+\| s\| \dot{u},
\end{equation}
where the dot denotes $\frac{d}{d\tau }$. Taking into account that $g\left( s,u\right) =0$ and (\ref{eqm22}), we have
\begin{equation}
\label{eqm23}
g\left( \dot{s},\frac{s}{\| s\| }\right) =g\left( u'\left( \tau '\right) \dot{\tau }'+\| s\| \dot{u},\frac{s}{\| s\| }\right) =\dot{\tau }'g\left( u'\left( \tau '\right) ,\frac{s}{\| s\| }\right) +g\left( \dot{u},s\right) ,
\end{equation}
and hence, by (\ref{eqm22}) and (\ref{eqm23}) we obtain
\begin{equation}
\label{eqm24}
\dot{s}=u'\left( \tau '\right) \dot{\tau }'+\left( \dot{\tau }'g\left( u'\left( \tau '\right) ,\frac{s}{\| s\| }\right) +g\left( \dot{u},s\right) -1\right) u+\| s\| \dot{u} .
\end{equation}
On the other hand
\begin{equation}
\label{eqm25}
g\left( s,u\right) =0 \,\Longrightarrow \, g\left( \dot{s},u\right) +g\left( s,\dot{u}\right) =0.
\end{equation}
Applying (\ref{eqm24}) in (\ref{eqm25}) and taking into account that $g\left( \dot{u},u\right) =0$, we find
\begin{equation}
\label{eqm26}
\dot{\tau }'=\frac{1}{g\left( u'\left( \tau '\right) ,\frac{s}{\| s\| }-u\right) }.
\end{equation}
Combining (\ref{eqm24}) and (\ref{eqm26}), we obtain
\begin{equation}
\label{fmink2}
\dot{s} = \frac{1}{g\left( u'\left( \tau '\right) ,\frac{s}{\| s\| }-u\right) }\left( u'\left( \tau '\right) +g\left( u'\left( \tau '\right) ,u\right) u\right) +g\left( s,\dot{u}\right) u+\| s\| \dot{u} .
\end{equation}

Let $U$, $U'$ be the 4-velocities of $\beta $ and $\beta
'$ respectively, and let $S=S_{\mathrm{obs}}$ (for the shake of readability) be the relative position of
$\beta '$ observed by $\beta $. Then, from
(\ref{fmink2}) we have
\begin{equation}
\label{fmink2_2}
\nabla _U S=\frac{1}{g\left( U',\frac{S}{\| S\| }-U\right) }\left( U'+g\left( U',U\right) U\right) +g\left( S,\nabla _U U\right) U+\| S\| \nabla _U U,
\end{equation}
where $U$, $S$, $\nabla _U U$, $\nabla _U S$ are evaluated at $p$, and $U'$ is evaluated at $q$.
So, by Proposition \ref{prop10} and expression (\ref{fmink2_2}),
the astrometric relative velocity $V_{\mathrm{ast}}$ of $\beta
'$ with respect to $\beta $ is given by
\begin{eqnarray*}
V_{\mathrm{ast}} &=&\nabla _U S-g\left(
S,\nabla _U U\right) U  \nonumber \\
&=&\frac{1}{g\left( U',\frac{S}{\| S\| }-U\right) }\left( U'+g\left( U',U\right) U\right) +\| S\| \nabla _U U,
\end{eqnarray*}
where $V_{\mathrm{ast}}$, $U$, $S$, $\nabla _U U$ are evaluated at $p$, and $U'$ is evaluated at $q$.
\end{proof}

Taking into account the expression of the spectroscopic relative velocity given in (\ref{vspec}), we obtain the next corollary:

\begin{corollary}
The astrometric relative velocity of $\beta '$ with respect to $\beta $ reads
\begin{equation}
\label{fmink2_3rel}
V_{\mathrm{ast}} =\| S_{\mathrm{obs}}\| \nabla _U U+\frac{1}{1+g\left( V_{\mathrm{spec}},\frac{S_{\mathrm{obs}}}{\| S_{\mathrm{obs}}\|}\right)}V_{\mathrm{spec}}.
\end{equation}
\end{corollary}

So, $V_{\mathrm{spec}}$ and $V_{\mathrm{ast}}$ are not
proportional unless $\beta $ is geodesic.

If $\beta ^{\prime }$ is geodesic then it is clear that
$V_{\mathrm{spec}}=V_{\mathrm{kin}}$. Moreover, if $\beta $ is
also geodesic then
$V_{\mathrm{spec}}=V_{\mathrm{kin}}=V_{\mathrm{Fermi}}$.

\begin{remark}
Let us suppose that $\beta $ and $\beta ^{\prime }$ intersect at
$p$, let $u$, $u^{\prime }$ be the 4-velocities of $\beta $,
$\beta ^{\prime }$ at $p$ respectively, and let $v$ be the
relative velocity of $u^{\prime }$ observed by $u$, in the sense
of expression (\ref{relvel}). Let us study the relations between
$v$, $V_{\mathrm{kin}~p}$, $V_{\mathrm{Fermi}~p}$,
$V_{\mathrm{spec}~p}$ and $V_{\mathrm{ast}~p} $.

It is clear that $V_{\mathrm{kin}~p}=V_{\mathrm{spec}~p}=v$, even
in general relativity. Moreover, since $S_{p}=0$, by
(\ref{fmink1_3}) we have $V_{\mathrm{Fermi}~p}=v$. On the other
hand, since $S_{\mathrm{obs}~p}=0$, it is easy to prove that
$V_{\mathrm{ast}~p}=\frac{1}{1\pm \left\Vert v\right\Vert }v$,
where we choose ``$+$" if we consider that $\beta ^{\prime }$ is
leaving from $\beta $, and we choose ``$-$" if we consider that
$\beta ^{\prime }$ is arriving at $\beta $. Therefore, if $\beta $
and $\beta ^{\prime }$ intersect at $p$, then it is not possible
to write $V_{\mathrm{ast}~p}$ in a unique way in terms of $v$.
\end{remark}

\begin{example}
\label{examink} Using rectangular coordinates $\left(
t,x,y,z\right) $, let us consider the following observers: $\beta \left( \tau \right)
:=\left( \tau ,0,0,0\right) $, and $\beta ^{\prime }\left( \tau
^{\prime }\right) :=\left\{
\begin{array}{l}
\left( \gamma \tau ^{\prime },v\gamma \tau ^{\prime },0,0\right)
~~{\mathrm{if}}~~\tau
^{\prime }\in \left[ 0,\frac{1}{\gamma v}\right] \\
\\
\left( \gamma \tau ^{\prime },2-v\gamma \tau ^{\prime },0,0\right)
~~{\mathrm{if}}~~\tau ^{\prime }\in \left] \frac{1}{\gamma
v},\frac{2}{\gamma v}\right]
\end{array}
\right. $ where $v\in \left] 0,1\right[ $ and $\gamma
:=\frac{1}{\sqrt{1-v^{2}}}$, parameterized by their proper times. That is, $\beta $ is a stationary
observer with $x=0$, $y=0$, $z=0$ and $\beta ^{\prime }$ is an
observer moving from $x=0$, $y=0$, $z=0$ to $x=1$, $y=0$, $z=0$
with velocity of modulus $v$ and returning (see Fig. \ref{fig1}).
It is satisfied that

\begin{figure}[tbp]
\begin{center}
\includegraphics[width=0.4\textwidth]{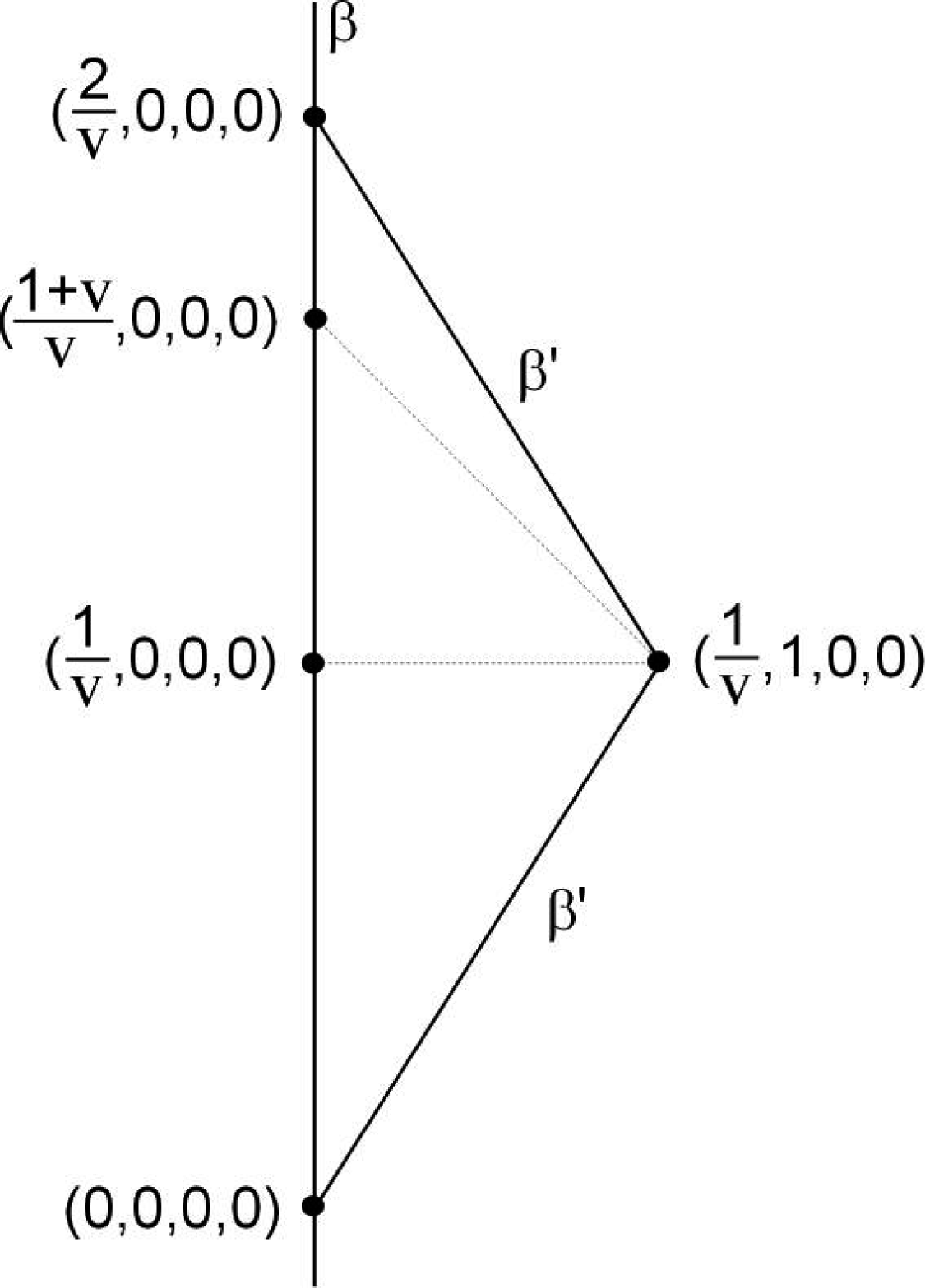}
\end{center}
\caption{Scheme of the observers of Example
\protect\ref{examink}.} \label{fig1}
\end{figure}

\[
V_{\mathrm{kin}~\beta \left( \tau \right) }=\left\{
\begin{array}{l}
v\left. \frac{\partial }{\partial x}\right\vert _{\beta \left(
\tau \right)
}~~\mathrm{if}~~\tau \in \left[ 0,\frac{1}{v}\right] \\
-v\left. \frac{\partial }{\partial x}\right\vert _{\beta \left(
\tau \right) }~~\mathrm{if}~~\tau \in \left]
\frac{1}{v},\frac{2}{v}\right]
\end{array}
\right. ,
\]
\[
V_{\mathrm{spec}~\beta \left( \tau \right) }=\left\{
\begin{array}{l}
v\left. \frac{\partial }{\partial x}\right\vert _{\beta \left(
\tau \right)
}~~\mathrm{if}~~\tau \in \left[ 0,\frac{1+v}{v}\right] \\
-v\left. \frac{\partial }{\partial x}\right\vert _{\beta \left(
\tau \right) }~~\mathrm{if}~~\tau \in \left]
\frac{1+v}{v},\frac{2}{v}\right]
\end{array}
\right. .
\]
Applying (\ref{fmink1_3}), we obtain $V_{\mathrm{Fermi}~\beta
\left( \tau \right) }=V_{\mathrm{kin}~\beta \left( \tau \right)
}$. Moreover
\[
S_{\mathrm{obs}~\beta \left( \tau \right) }=\left\{
\begin{array}{l}
\frac{v\tau }{1+v}\left. \frac{\partial }{\partial x}\right\vert
_{\beta
\left( \tau \right) }~~\mathrm{if}~~\tau \in \left[ 0,\frac{1+v}{v}\right] \\
\frac{2-v\tau }{1-v}\left. \frac{\partial }{\partial x}\right\vert
_{\beta \left( \tau \right) }~~\mathrm{if}~~\tau \in \left]
\frac{1+v}{v},\frac{2}{v}\right]
\end{array}
\right. .
\]
Hence, by (\ref{fmink2_3}) we have
\[
V_{\mathrm{ast}~\beta \left( \tau \right) }=\left\{
\begin{array}{l}
\frac{v}{1+v}\left. \frac{\partial }{\partial x}\right\vert
_{\beta \left(
\tau \right) }~~\mathrm{if}~~\tau \in \left[ 0,\frac{1+v}{v}\right] \\
-\frac{v}{1-v}\left. \frac{\partial }{\partial x}\right\vert
_{\beta \left( \tau \right) }~~\mathrm{if}~~\tau \in \left]
\frac{1+v}{v},\frac{2}{v}\right]
\end{array}
\right. .
\]
Consequently, $\left\Vert V_{\mathrm{ast}~\beta \left( \tau
\right) }\right\Vert \in \left] 0,1/2\right[ $ if $\tau \in \left[
0,\frac{1+v}{v}\right] $, i.e. if $\beta ^{\prime }$ is moving
away radially. On the other hand, $\left\Vert
V_{\mathrm{ast}~\beta \left( \tau \right) }\right\Vert \in \left]
0,+\infty \right[ $ if $\tau \in \left]
\frac{1+v}{v},\frac{2}{v}\right] $, i.e. if $\beta ^{\prime }$ is
getting closer radially (see fig. \ref{newfig1}). This corresponds to what $\beta $
observes.

\begin{figure}[tbp]
\begin{center}
\includegraphics[width=0.4\textwidth]{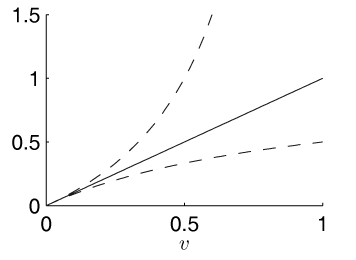}
\end{center}
\caption{Modulus of the relative velocities of Example
\protect\ref{examink} depending on the parameter $v$. The solid line represents the modulus of $V_{\mathrm{kin}}$, $V_{\mathrm{Fermi}}$ and $V_{\mathrm{spec}}$, and they are always equal to $v$. The dashed line represents the modulus of $V_{\mathrm{ast}}$ when $\beta '$ moves away from $\beta $ (lower) and $\beta '$ approaches $\beta $ (upper).} \label{newfig1}
\end{figure}

\end{example}

\begin{example}
Let us suppose that the spacetime is flat and we see an alien
spaceship coming to Earth from a planet at 9 lightyears
(this distance can be measured by parallax, since this method
estimates the affine distance from the planet to Earth observed
by someone on Earth). Let us suppose that the spaceship is coming
radially, and so, we can measure the modulus of its spectroscopic
relative velocity (see \ref{rem2}). Supposing that this modulus is
$v=0.9$, the spaceship will take 10 years to arrive at Earth from
its planet. However, since light takes 9 years to arrive at us,
there is only 1 year left for the arrival of the spaceship. This
result can also be obtained by using expression (\ref{fmink2_3}):
in our case, the modulus of the astrometric relative velocity is
$\frac{0.9}{1-0.9}=9$, and we will therefore observe that it takes
1 year to arrive.
\end{example}

\begin{remark}
There is an open problem in general relativity, that consists on finding expressions for $V_{\mathrm{Fermi}}$ and $V_{\mathrm{ast}}$ in terms of
$U$, $\nabla _U U$, $U'$, $S$ and $S_{\mathrm{obs}}$, analogously to Propositions \ref{propmink1} and \ref{propmink2}, avoiding $\nabla _U S$, $\nabla _U S_{\mathrm{obs}}$, or any term involving the evolution of $S$, $S_{\mathrm{obs}}$. It would be very useful in the calculations of the relative velocities.
\end{remark}

\section{Examples in General Relativity}
\label{sec6}

In this section, we are going to study some fundamental examples in Schwarzschild and Robertson-Walker spacetimes. See \cite{Klein10} for an interesting and complete study
of the relative velocities of a radially receding test particle with respect to / observed by a central observer in a Schwarzschild-de Sitter spacetime.

\subsection{Stationary observers in Schwarzschild spacetime}

In the Schwarzschild metric with spherical coordinates
\[
\mathrm{d}s^{2}=-a^{2}\left( r\right)
\mathrm{d}t^{2}+\frac{1}{a^{2}\left( r\right)
}\mathrm{d}r^{2}+r^{2}\left( \mathrm{d}\theta ^{2}+\sin ^{2}\theta
\mathrm{d}\varphi ^{2}\right) ,
\]
where $a\left( r\right) =\sqrt{1-\frac{2m}{r}}$ and $r>2m$, let us
consider two equatorial stationary observers, $\beta _{1}\left(
\tau \right) =\left( \frac{1}{a_{1}}\tau ,r_{1},\pi /2,0\right) $
and $\beta _{2}\left( \tau \right) =\left( \frac{1}{a_{2}}\tau
,r_{2},\pi /2,0\right) $ with $\tau \in \mathbb{R}$,
$r_{2}>r_{1}>2m$, $a_1:=a\left( r_1\right) $ and $a_2:=a\left(
r_2\right) $, and let $U$ be the 4-velocity of $\beta _{2}$, i.e.
$U:=\frac{1}{a_{2}}\frac{\partial }{\partial t}$. We are going to
study the relative velocities of $\beta _{1}$ with respect to / observed by $\beta _{2}$.

\subsubsection{Kinematic and Fermi relative velocities. Fermi distance}

Let us consider the vector field $X:=a\left( r\right)
\frac{\partial }{\partial r}$; it is spacelike,
unit, geodesic, and orthogonal to $U$. Since $\nabla _{X}\left(
\frac{1}{a\left( r\right) }\frac{\partial }{\partial t}\right)
=0$, we have that the kinematic relative velocity
$V_{\mathrm{kin}} $ of $\beta _{1}$ with respect to $\beta _{2}$
is given by $V_{\mathrm{kin}}=0$.

It is clear (a priori) that the relative position $S$ of $\beta _{1}$ with
respect to $\beta _{2}$ is proportional to $\frac{\partial }{\partial r}$ and the proportionality factor is constant.
So, it is easy to prove that $\nabla _{U}S$ is proportional to $U$ and therefore, the Fermi relative velocity $V_{\mathrm{Fermi}}$ of
$\beta _{1}$ with respect to $\beta _{2}$ reads
$V_{\mathrm{Fermi}}=0$.

Nevertheless, we are going to calculate the Fermi distance and $S$:

Let $\alpha \left( \sigma \right) =\left( t_{0},\alpha ^{r}\left(
\sigma \right) ,\pi /2,0\right) $ be an integral curve of $X$ such
that $q:=\alpha \left( \sigma _{1}\right) \in \beta _{1}$ and
$p:=\alpha \left( \sigma _{2}\right) \in \beta _{2}$, with $\sigma
_{2}>\sigma _{1}$ (i.e. $\alpha \left( \sigma \right) $ is a
spacelike geodesic from $q$ to $p$, parameterized by its
arclength, and its tangent vector at $p$ is $X_{p}$). Then, by
Proposition \ref{propdspaceparam}, the Fermi distance
$d_{U_{p}}^{\mathrm{Fermi}}\left( q,p\right) $ from $q$ to $p$
with respect to $U_{p}$ is $\sigma _{2}-\sigma _{1}$. Since
$\alpha $ is an integral curve of $X$, we have $\overset{.}{\alpha
}^{r}\left( \sigma \right) =\sqrt{1-\frac{2m}{\alpha ^{r}\left(
\sigma \right) }}$. So, $\int_{r_{1}}^{r_{2}}\left(
1-\frac{2m}{\alpha ^{r}\left( \sigma \right) }\right)
^{-1/2}\overset{.}{\alpha }^{r}\left( \sigma \right)
\mathrm{d}\sigma =\sigma _{2}-\sigma _{1}$, and then
\begin{equation}
d_{U_{p}}^{\mathrm{Fermi}}\left( q,p\right) =2m\ln \left(
\frac{\left( 1-a_{1}\right) \sqrt{r_{1}}}{\left( 1-a_{2}\right)
\sqrt{r_{2}}}\right) +r_{2}a_{2}-r_{1}a_{1}.  \label{dspaceschw}
\end{equation}%
Since (\ref{dspaceschw}) does not depends on $t_{0}$, the Fermi
distance from $\beta _{1}$ to $\beta _{2}$ with respect to $\beta
_{2}$ is also given by expression (\ref{dspaceschw}). Hence, by
(\ref{duspaceqp}), the relative position $S$ of $\beta _{1}$ with
respect to $\beta _{2}$ is given by
\[
S=\left( 2m\ln \left( \frac{\left( 1-a_{2}\right)
\sqrt{r_{2}}}{\left( 1-a_{1}\right) \sqrt{r_{1}}}\right)
+r_{1}a_{1}-r_{2}a_{2}\right) a_{2}\frac{\partial }{\partial r}.
\]

\subsubsection{Spectroscopic and astrometric relative velocities. Affine distance}

\begin{figure}[tbp]
\begin{center}
\includegraphics[width=0.8\textwidth]{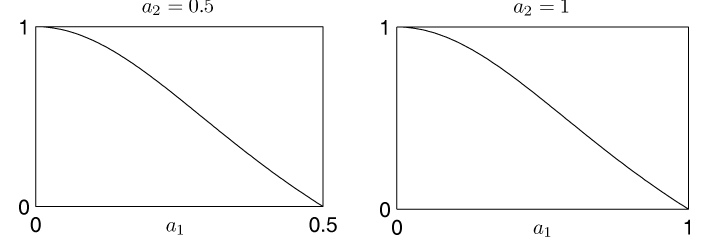}
\end{center}
\caption{Modulus of $V_{\mathrm{spec}}$ of a stationary observer with $a_1=\sqrt{1-\frac{2m}{r_1}}$ observed by another stationary observer with $a_2=\sqrt{1-\frac{2m}{r_2}}=0.5$ (left) and at the exterior limit $a_2=1$ ($r_2=+\infty $) (right) in Schwarzschild spacetime. It produces the \textit{gravitational redshift}.} \label{newfig2}
\end{figure}

It is easy to prove that the spectroscopic relative velocity
$V_{\mathrm{spec}}$ of $\beta _{1}$ observed by $\beta _{2}$ is
radial. Since the gravitational redshift is given by
$\frac{a_{2}}{a_{1}}$ (see \cite{Bolo05}), by (\ref{vspecshift})
we obtain
\begin{equation}
V_{\mathrm{spec}}=-a_{2}\frac{a_{2}^{2}-a_{1}^{2}}{a_{2}^{2}+a_{1}^{2}}\frac{\partial
}{\partial r}.  \label{vspeca2a12}
\end{equation}
Expression (\ref{vspeca2a12}) is also obtained in \cite{Bolo05}.
Since $\left\Vert
V_{\mathrm{spec}}\right\Vert =\frac{a_{2}^{2}-a_{1}^{2}}{a_{2}^{2}+a_{1}^{2}}$, we have $\lim_{r_{1}\rightarrow 2m}\left\Vert
V_{\mathrm{spec}}\right\Vert =1$ (see Fig. \ref{newfig2}).

On the other hand, it is clear (a priori) that the relative position $S_{\mathrm{obs}}$ of $\beta _{1}$ observed by
$\beta _{2}$ is proportional to $\frac{\partial }{\partial r}$ and the proportionality factor is constant.
So, it is easy to prove that $\nabla _{U}S_{\mathrm{obs}}$ is proportional to $U$ and therefore, the astrometric relative velocity $V_{\mathrm{ast}}$ of
$\beta _{1}$ observed by $\beta _{2}$ reads
$V_{\mathrm{ast}}=0$.

Nevertheless, we are going to calculate the affine distance and $S_{\mathrm{obs}}$:

In \cite{Bolo05} it is proved (by using
Proposition \ref{propdlightparam}) that the affine distance from
$\beta _{1}$ to $\beta _{2}$ observed by $\beta _{2}$ is
$\frac{r_{2}-r_{1}}{a_{2}}$. Hence, by (\ref{dulightqp}), the
relative position $S_{\mathrm{obs}}$ of $\beta _{1}$ observed by
$\beta _{2}$ is given by
\begin{equation}
S_{\mathrm{obs}}=\left( r_{1}-r_{2}\right) \frac{\partial
}{\partial r}. \label{sobsr1r2}
\end{equation}

\subsection{Free-falling observers in Schwarzschild spacetime}
\label{free-falling}

Let us consider a radial free-falling observer $\beta _{1}$
parameterized by the coordinate time $t$, $\beta _{1}\left(
t\right) =\left( t,\beta _{1}^{r}\left( t\right) ,\pi /2,0\right)
$. Given an event $q=\left( t_{1},r_{1},\pi /2,0\right) \in \beta
_{1}$, the 4-velocity of $\beta _{1}$ at $q$ is given by
\begin{equation}
u_{1}=\frac{E}{a_{1}^{2}}\left. \frac{\partial }{\partial
t}\right\vert _{q}-\sqrt{E^{2}-a_{1}^{2}}\left. \frac{\partial
}{\partial r}\right\vert _{q}, \label{4velfreefall}
\end{equation}
where $E$ is a constant of motion given by $E:=\left(
\frac{1-2m/r_{0}}{1-v_{0}^{2}}\right) ^{1/2}$, $r_{0}$ is the
radial coordinate at which the fall begins, $v_{0}$ is the initial
velocity (see \cite{Craw02}), and $a_{1}:=a\left( r_{1}\right) $.
Moreover, let us consider an equatorial stationary observer $\beta
_{2}\left( \tau \right) =\left( \frac{1}{a_{2}}\tau ,r_{2},\pi
/2,0\right) $ with $\tau \in \mathbb{R}$, $r_{2}\geq r_{1}>2m$,
$a_{2}:=a\left( r_{2}\right) $, and
$U:=\frac{1}{a_{2}}\frac{\partial }{\partial t}$ its 4-velocity.
We are going to study the relative velocities of $\beta _{1}$ with
respect to / observed by $\beta _{2}$ at $p$, where $p$ will be
a determined event of $\beta _{2}$.

\subsubsection{Kinematic and Fermi relative velocities}

Let $p=\left( t_{1},r_{2},\pi /2,0\right) $. This is the unique
event of $\beta _{2}$ such that $q\in L_{p,U_{p}}$, i.e. there
exists a spacelike geodesic $\alpha \left( \sigma \right) $ from
$q=\alpha \left( \sigma _{1}\right) $ to $p=\alpha \left( \sigma
_{2}\right) $ such that the tangent vector $\overset{.}{\alpha
}\left( \sigma _{2}\right) $ is orthogonal to $U_{p}$. We can
consider $\alpha \left( \sigma \right) $ parameterized by its
arclength and $\sigma _{2}>\sigma _{1}$. So, $\alpha \left( \sigma
\right) $ is an integral curve of the vector field $X=a\left(
r\right) \frac{\partial }{\partial r}$. If we parallelly transport
$u_{1}$ from $q$ to $p$ along $\alpha $ we obtain $\tau
_{qp}u_{1}=\frac{E}{a_{1}a_{2}}\left. \frac{\partial }{\partial
t}\right\vert _{p}-\frac{a_{2}}{a_{1}}\sqrt{E^{2}-a_{1}^{2}}\left.
\frac{\partial }{\partial r}\right\vert _{p}$. By
(\ref{fkinrelvel}), the kinematic relative velocity
$V_{\mathrm{kin}~p}$ of $\beta _{1}$ with respect to $\beta _{2}$
at $p$ reads
\[
V_{\mathrm{kin}~p}=-a_{2}\sqrt{1-\frac{a_1^2}{E^2}}\left.
\frac{\partial }{\partial r}\right\vert _{p}.
\]
Since $\left\Vert V_{\mathrm{kin}~p}\right\Vert =\sqrt{1-\frac{a_1^2}{E^2}}$, it is satisfied that $\lim_{r_{1}\rightarrow 2m}\left\Vert
V_{\mathrm{kin}~p}\right\Vert =1$. See Appendix for a deeper analysis of this function.

On the other hand, by (\ref{dspaceschw}), the relative position
$S$ of $\beta _{1}$ with respect to $\beta _{2}$ is given by
\[
S=\left( 2m\ln \left( \frac{\left( 1-a_{2}\right)
\sqrt{r_{2}}}{\left( 1-a\left( \beta _{1}^{r}\left( t\right)
\right) \right) \sqrt{\beta _{1}^{r}\left( t\right) }}\right)
+\beta _{1}^{r}\left( t\right) a\left( \beta _{1}^{r}\left(
t\right) \right) -r_{2}a_{2}\right) a_{2}\frac{\partial }{\partial
r}.
\]
By (\ref{defvFermiusgusuu}), the Fermi relative velocity
$V_{\mathrm{Fermi}}$ of $\beta _{1}$ with respect to $\beta _{2}$
reads
\[
V_{\mathrm{Fermi}}=\left( \nabla _{U}S\right) ^{r}\frac{\partial
}{\partial r}=\frac{1}{a_{2}}\frac{\partial S^{r}}{\partial
t}\frac{\partial }{\partial
r}=\frac{1}{a_{2}}\frac{\overset{.}{\beta }_{1}^{r}\left( t\right)
}{a\left( \beta _{1}^{r}\left( t\right) \right) }\frac{\partial
}{\partial r}
\]
Taking into account (\ref{4velfreefall}), we have
$\overset{.}{\beta }_{1}^{r}\left( t_{1}\right)
=-a_{1}^{2}\sqrt{1-\frac{a_1^2}{E^2}}$. Hence
\[
V_{\mathrm{Fermi}~p}=-\frac{a_{1}}{a_{2}}\sqrt{1-\frac{a_1^2}{E^2}}\left.
\frac{\partial }{\partial r}\right\vert _{p}.
\]
Since $\left\Vert
V_{\mathrm{Fermi}~p}\right\Vert =\frac{a_1}{a^2 _2}\sqrt{1-\frac{a_1^2}{E^2}}$, it is satisfied that $\lim_{r_{1}\rightarrow 2m}\left\Vert
V_{\mathrm{Fermi}~p}\right\Vert =0$. See Appendix for a deeper analysis of this function.

\subsubsection{Spectroscopic and astrometric relative velocities}

Let $p$ be the unique event of $\beta _{2}$ such that there exists
a light ray $\lambda $ from $q$ to $p$, and let us suppose that
$p=\left( t_{2},r_{2},\pi /2,0\right) $. In \cite{Bolo05} it is
shown that the spectroscopic relative velocity
$V_{\mathrm{spec}~p}$ of $\beta _{1}$ observed by $\beta _{2}$ at
$p$ is given by
\begin{equation}
\label{vspecff} V_{\mathrm{spec}~p}=-a_{2}\frac{\left(
a_{2}^{2}+a_{1}^{2}\right) \sqrt{1-\frac{a_1^2}{E^2}}+\left(
a_{2}^{2}-a_{1}^{2}\right) }{\left( a_{2}^{2}-a_{1}^{2}\right)
\sqrt{1-\frac{a_1^2}{E^2}}+\left( a_{2}^{2}+a_{1}^{2}\right) }\left.
\frac{\partial }{\partial r}\right\vert _{p}.
\end{equation}
Since $\left\Vert V_{\mathrm{spec}~p}\right\Vert =\frac{\left(
a_{2}^{2}+a_{1}^{2}\right) \sqrt{1-\frac{a_1^2}{E^2}}+\left(
a_{2}^{2}-a_{1}^{2}\right) }{\left( a_{2}^{2}-a_{1}^{2}\right)
\sqrt{1-\frac{a_1^2}{E^2}}+\left( a_{2}^{2}+a_{1}^{2}\right) }$, it follows that $\lim_{r_{1}\rightarrow 2m}\left\Vert
V_{\mathrm{spec}~p}\right\Vert =1$. See Appendix for a deeper analysis of this function.

On the other hand, it can be checked that
\[
\lambda \left( r\right) :=\left( t_{1}+r-r_{1}+2m\ln \left(
\frac{r-2m}{r_{1}-2m}\right) ,r,\pi /2,0\right) ,\quad r\in
[r_1,r_2]
\]
is a light ray from $q=\lambda \left( r_{1}\right) $ to $p=\lambda
\left( r_{2}\right) $. So,
\begin{equation}
t_{2}=\lambda ^{t}\left( r_{2}\right) =t_{1}+r_{2}-r_{1}+2m\ln
\left( \frac{r_{2}-2m}{r_{1}-2m}\right) .  \label{t2lambdatr2}
\end{equation}
Let us define implicitly the function $f\left( t\right) $ by the
expression
\begin{equation}
f\left( t\right) :=t-\left( r_{2}-\beta _{1}^{r}\left( f\left(
t\right) \right) +2m\ln \left( \frac{r_{2}-2m}{\beta
_{1}^{r}\left( f\left( t\right) \right) -2m}\right) \right) .
\label{deftmenos}
\end{equation}
Taking into account (\ref{t2lambdatr2}), $f\left( t\right) $ is
the coordinate time at which $\beta _{1}$ emits a light ray that
arrives at $\beta _{2}$ at coordinate time $t$. Applying
(\ref{sobsr1r2}), the relative position $S_{\mathrm{obs}}$ of
$\beta _{1}$ observed by $\beta _{2}$ reads
\[
S_{\mathrm{obs}}=\left( \beta _{1}^{r}\left( f\left( t\right)
\right) -r_{2}\right) \frac{\partial }{\partial r}.
\]
By (\ref{defvastusgusuu}), the astrometric relative velocity
$V_{\mathrm{ast}}$ of $\beta _{1}$ observed by $\beta _{2}$ is
given by
\[
V_{\mathrm{ast}}=\left( \nabla _{U}S_{\mathrm{obs}}\right)
^{r}\frac{\partial }{\partial r}=\frac{1}{a_{2}}\frac{\partial
S_{\mathrm{obs}}^{r}}{\partial t}\frac{\partial }{\partial
r}=\frac{1}{a_{2}}\overset{.}{\beta }_{1}^{r}\left( f\left(
t\right) \right) \overset{.}{f}\left( t\right) \frac{\partial
}{\partial r}.
\]
From (\ref{deftmenos}), we have $\overset{.}{f}\left( t_{2}\right)
=\frac{a_{1}^{2}}{a_{1}^{2}-\left( a_{1}^{2}-1\right)
\overset{.}{\beta }_{1}^{r}\left( t_{1}\right) }$. Moreover,
taking into account (\ref{4velfreefall}), we have
$\overset{.}{\beta }_{1}^{r}\left( t_{1}\right)
=-a_{1}^{2}\sqrt{1-\frac{a_1^2}{E^2}}$. Hence
\begin{equation}
\label{vastff}
V_{\mathrm{ast}~p}=-\frac{a_{1}^{2}}{a_{2}}\frac{\sqrt{1-\frac{a_1^2}{E^2}}}{1+\left(
a_{1}^{2}-1\right) \sqrt{1-\frac{a_1^2}{E^2}}}\left. \frac{\partial
}{\partial r}\right\vert _{p},
\end{equation}
and, in consequence, $\left\Vert V_{\mathrm{ast}~p}\right\Vert
=\frac{a_{1}^{2}}{a^2_{2}}\frac{\sqrt{1-\frac{a_1^2}{E^2}}}{1+\left(
a_{1}^{2}-1\right) \sqrt{1-\frac{a_1^2}{E^2}}}$, concluding that $\lim_{r_{1}\rightarrow 2m}\left\Vert
V_{\mathrm{ast}~p}\right\Vert
=\frac{1}{a_{2}^{2}}\frac{2E^{2}}{1+2E^{2}}\in \left] 0,+\infty
\right[ $. See Appendix for a deeper analysis of this function.

\subsection{Comoving observers in Robertson-Walker spacetime}

See \cite{Klein11} for an interesting and complete study
of the Fermi relative velocity of a comoving test particle with respect to / observed by a comoving observer in an expanding Robertson-Walker spacetime.
Currently, there is a work in process that studies the other relative velocities in this case, with examples in the Milne,
de-Sitter, radiation-dominated an matter-dominated universes.

In a Robertson-Walker metric with cartesian coordinates
\[
\mathrm{d}s^{2}=-\mathrm{d}t^{2}+\frac{a^{2}\left( t\right)
}{\left( 1+\frac{1}{4}kr^{2}\right) ^{2}}\left(
\mathrm{d}x^{2}+\mathrm{d}y^{2}+\mathrm{d}z^{2}\right) ,
\]
where $a\left( t\right) $ is the scale factor, $k=-1,0,1$ and
$r:=\sqrt{x^{2}+y^{2}+z^{2}}$, we consider two comoving (in the
classical sense, see \cite{Sach77}) observers $\beta _{0}\left(
\tau \right) =\left( \tau ,0,0,0\right) $ and $\beta _{1}\left(
\tau \right) =\left( \tau ,x_{1},0,0\right) $ with $\tau \in
\mathbb{R}$ and $x_{1}>0$. Let $t_{0}\in \mathbb{R}$, $p:=\beta
_{0}\left( t_{0}\right) $ and $u:=\overset{.}{\beta _{0}}\left(
t_{0}\right) =\left. \frac{\partial }{\partial t}\right\vert _{p}$
(i.e. the 4-velocity of $\beta _{0}$ at $p$). We are going to
study the relative velocities of $\beta _{1}$ with respect to / observed by $\beta _{0}$ at $p$.

\subsubsection{Kinematic and Fermi relative velocities}

The vector field $$X:=-\sqrt{\frac{a_{0}^{2}}{a^{2}\left( t\right)
}-1}\frac{\partial }{\partial t}+\frac{a_{0}}{a^{2}\left( t\right)
}\left( 1+\frac{1}{4}kx^{2}\right) \frac{\partial }{\partial x}$$
is geodesic, spacelike, unit, and $X_{p}$ is orthogonal to $u$,
i.e. it is tangent to the Landau submanifold $L_{p,u}$. Let $\beta
_{1}\left( t_{1}\right) =:q$ be the unique event of $\beta
_{1}\cap L_{p,u}$. We can find $t_{1}$ for a given scale factor
$a\left( t\right) $ taking into account the expression of $X$, but
we can not find an explicit expression in the general case. If
$u^{\prime }:=\overset{.}{\beta _{1}}\left( t_{1}\right) =\left.
\frac{\partial }{\partial t}\right\vert _{q}$, then $\tau
_{qp}u^{\prime }=\frac{a_{0}}{a_{1}}\left. \frac{\partial
}{\partial t}\right\vert
_{p}+\sqrt{\frac{1}{a_{1}^{2}}-\frac{1}{a_{0}^{2}}}\left.
\frac{\partial }{\partial x}\right\vert _{p}$, where
$a_{1}:=a\left( t_{1}\right) $ (it is well defined because
$a_{0}\geq a_{1}>0$). So, by (\ref{fkinrelvel}), the kinematic
relative velocity $V_{\mathrm{kin}~p}$ of $\beta _{1}$ with
respect to $\beta _{0}$ at $p$ is given by
\[
V_{\mathrm{kin}~p}=\frac{1}{a_{0}^{2}}\sqrt{a_{0}^{2}-a_{1}^{2}}\left.
\frac{\partial }{\partial x}\right\vert _{p}.
\]

Given a scale factor $a\left( t\right) $, the Fermi
distance $d^{\mathrm{Fermi}}$ from $\beta _{1}$ to $\beta _{0}$
with respect to $\beta _{0}$ can be also found, taking into
account the expression of $X$. So, the relative position $S$ of
$\beta _{1}$ with respect to $\beta _{0}$ reads
\[
S=d^{\mathrm{Fermi}}\frac{\left( 1+\frac{1}{4}kr^{2}\right)
}{a\left( t\right) }\frac{\partial }{\partial x},
\]
because $d^{\mathrm{Fermi}}=\left\Vert S\right\Vert $. Hence, the
Fermi relative velocity $V_{\mathrm{Fermi}~p}$ of $\beta _{1}$
with respect to $\beta _{0}$ at $p$ is given by
\[
V_{\mathrm{Fermi}~p}=\left( \left.
\frac{\mathrm{d}}{\mathrm{d}t}\left(
\frac{d^{\mathrm{Fermi}}}{a\left( t\right) }\right) \right\vert
_{t=t_{0}}+d_{p}^{\mathrm{Fermi}}\frac{\overset{.}{a}\left(
t_{0}\right) }{a_{0}^{2}}\right) \left. \frac{\partial }{\partial
x}\right\vert _{p}.
\]

\subsubsection{Spectroscopic and astrometric relative velocities}

Let $\lambda $ be a light ray received by $\beta _{0}$ at $p$ and
emitted from $\beta _{1}$ at $\beta _{1}\left( t_{1}\right) $.
Note that $t_{1}$ can be found from $x_{1}$ and $t_{0}$ taking
into account that
$\int_{0}^{x_{1}}\frac{\mathrm{d}x}{1+\frac{1}{4}kx^{2}}=\int_{t_{1}}^{t_{0}}\frac{\mathrm{d}t}{a\left(
t\right) }$. It can be easily proved that the spectroscopic
relative velocity $V_{\mathrm{spec}~p}$ of $\beta _{1}$ observed
by $\beta _{0}$ at $p$ is radial (by isotropy). So, by
(\ref{vspecshift}) taking into account that the cosmological shift
is given by $\frac{a_{0}}{a_{1}}$ (see \cite{Bolo05}), where
$a_{0}:=a\left( t_{0}\right) $ and $a_{1}:=a\left( t_{1}\right) $,
we have
\begin{equation}
V_{\mathrm{spec}~p}=\frac{1}{a_{0}}\frac{a_{0}^{2}-a_{1}^{2}}{a_{0}^{2}+a_{1}^{2}}\left.
\frac{\partial }{\partial x}\right\vert _{p}. \label{vspecprob}
\end{equation}

Given a scale factor $a\left( t\right) $, the affine
distance $d^{\mathrm{aff}}$ from $\beta _{1}$ to $\beta _{0}$
observed by $\beta _{0}$ can be found. So, the relative position
$S_{\mathrm{obs}}$ of $\beta _{1}$ observed by $\beta _{0}$ is
given by
\[
S_{\mathrm{obs}}=d^{\mathrm{aff}}\frac{\left(
1+\frac{1}{4}kr^{2}\right) }{a\left( t\right) }\frac{\partial
}{\partial x},
\]
because $d^{\mathrm{aff}}=\left\Vert
S_{\mathrm{obs}}\right\Vert $. Hence, the astrometric relative
velocity $V_{\mathrm{ast}~p}$ of $\beta _{1}$ observed by $\beta
_{0}$ at $p$ reads
\begin{equation}
V_{\mathrm{ast}~p}=\left( \left.
\frac{\mathrm{d}}{\mathrm{d}t}\left(
\frac{d^{\mathrm{aff}}}{a\left( t\right) }\right) \right\vert
_{t=t_{0}}+d_{p}^{\mathrm{aff}}\frac{\overset{.}{a}\left(
t_{0}\right) }{a_{0}^{2}}\right) \left. \frac{\partial }{\partial
x}\right\vert _{p}.  \label{vastprob}
\end{equation}

Let us study these relative velocities in more detail. In
cosmology it is usual to consider the scale factor in the form
\[
a\left( t\right) =a_{0}\left( 1+H_{0}\left( t-t_{0}\right)
-\frac{1}{2}q_{0}H_{0}^{2}\left( t-t_{0}\right) ^{2}\right)
+\mathcal{O}\left( H_{0}^{3}\left( t-t_{0}\right) ^{3}\right) ,
\]
where $t_{0}\in \mathbb{R}$, $a_{0}=a\left( t_{0}\right) >0$,
$H\left( t\right) =\overset{.}{a}\left( t\right) /a\left( t\right)
$ is the Hubble \textquotedblleft constant\textquotedblright ,
$H_{0}=H\left( t_{0}\right) >0$, $q\left( t\right) =-a\left(
t\right) \overset{..}{a}\left( t\right) /\overset{.}{a}\left(
t\right) ^{2}$ is the deceleration coefficient, and $q_{0}=q\left(
t_{0}\right) $, with $\left\vert H_{0}\left( t-t_{0}\right)
\right\vert \ll 1$ (see \cite{Misn73}). This corresponds to a
universe in decelerated expansion and the time scales that we are
going to use are relatively small. Let us define $p:=\beta
_{0}\left( t_{0}\right) $ and $u:=\overset{.}{\beta _{0}}\left(
t_{0}\right) =\left. \frac{\partial }{\partial t}\right\vert _{p}
$.

We are going to express the spectroscopic and the astrometric
relative velocity of $\beta _{1}$ observed by $\beta _{0}$ at $p$
in terms of the redshift parameter at $t=t_{0}$, defined as
$z_{0}:=\frac{a_{0}}{a_{1}}-1$, where $a_{1}:=a\left( t_{1}\right)
$. This parameter is very usual in cosmology since it can be
measured by spectroscopic observations. By (\ref{vspecprob}), the
spectroscopic relative velocity $V_{\mathrm{spec}~p}$ of $\beta
_{1} $ observed by $\beta _{0}$ at $p$ is given by
\begin{equation}\label{vspecrw}
V_{\mathrm{spec}~p}=\frac{1}{a_{0}}\frac{a_{0}^{4}-\left(
z_{0}+1\right) ^{2}}{a_{0}^{4}+\left( z_{0}+1\right) ^{2}}\left.
\frac{\partial }{\partial x}\right\vert _{p}.
\end{equation}

In \cite{Bolo05} it is shown that the affine distance
$d^{\mathrm{aff}}$ from $\beta _{1}$ to $\beta _{0}$ observed
by $\beta _{0}$ reads
\[
d^{\mathrm{aff}}\left( t\right) =\frac{z\left( t\right)
}{H\left( t\right) }\left( 1-\frac{1}{2}\left( 3+q\left( t\right)
\right) z\left( t\right) \right) +\mathcal{O}\left( z^{3}\left(
t\right) \right) ,
\]
where $z\left( t\right) $ is the redshift function. So, by
(\ref{vastprob}), the astrometric relative velocity
$V_{\mathrm{ast}~p}$ of $\beta _{1}$ observed by $\beta _{0}$ at
$p$ is given by
\[
V_{\mathrm{ast}~p}=\left( \frac{\overset{.}{z}\left( t_{0}\right)
}{a_{0}H_{0}}+\frac{z_{0}}{a_{0}}\left(
q_{0}+1-\frac{\overset{.}{z}\left( t_{0}\right) }{H_{0}}\left(
3+q_{0}\right) \right) +\mathcal{O}\left( z_{0}^{2}\right) \right)
\left. \frac{\partial }{\partial x}\right\vert _{p}.
\]%
Hence, if we suppose that $\overset{.}{z}\left( t_{0}\right)
\approx 0$ (i.e., the redshift is constant in our time scale),
then
\begin{equation}
\label{vastrw} V_{\mathrm{ast}~p}\approx \left(
\frac{z_{0}}{a_{0}}\left( q_{0}+1\right) +\mathcal{O}\left(
z_{0}^{2}\right) \right) \left. \frac{\partial }{\partial
x}\right\vert _{p}.
\end{equation}

\section{Discussion and comments}

It is usual to consider the spectroscopic relative velocity as a
non-acceptable ``physical velocity''. However, in this paper we
have defined it in a geometric way, showing that it is, in fact, a
very plausible physical velocity.
\begin{itemize}

\item Firstly, in other works (see \cite{Bolo02}, \cite{Bolo05}),
we have discussed pros and cons of spacelike and lightlike
simultaneities, coming to the conclusion that lightlike
simultaneity is physically and mathematically more suitable. Since
the spectroscopic relative velocity is the natural generalization
(in the framework of lightlike simultaneity) of the usual concept
of relative velocity (given by (\ref{relvel})), it might have a
lot of importance.

\item Secondly, there are some good properties suggesting that the
spectroscopic relative velocity has a lot of physical sense. For
instance, if we work with the spectroscopic relative velocity, it
is shown in \cite{Bolo05} that gravitational redshift is just a
particular case of a generalized Doppler effect.

\end{itemize}

Nevertheless, all four concepts of relative velocity have full
physical sense and they must be studied equally.

Finally, one can wonder whether the discussed concepts of relative
velocity can be actually determined experimentally. A priori, only
the spectroscopic and astrometric relative velocities can be
measured by direct observation. The shift allows us to find
relations between the modulus of the spectroscopic relative
velocity and its tangential component, as we show in
(\ref{fshiftvrad}). But, in general, it is not enough information
to determine it completely (as we discuss in Remark
\ref{remark1}), unless we make some assumptions (see Remark
\ref{rem2}) or we use a model for the spacetime and apply some
expressions like (\ref{vspeca2a12}), (\ref{vspecff}), or
(\ref{vspecrw}). Finding the astrometric relative velocity is
basically the same problem as finding the optical coordinates. It
is non-trivial and it has been widely treated, for instance, in
\cite{Elli85}. Nevertheless, expressions like (\ref{vastff}) or
(\ref{vastrw}) could be very useful in particular situations.
Since the measure of these velocities is rather difficult, any
expression relating them can be very helpful in order to determine
them, as, for example, expression (\ref{fmink2_3}) in special
relativity.

\section*{Appendix}

\subsection*{A.1 Free-falling observers in Schwarzschild spacetime}

We are going to study the modulus of the relative velocities of a radially inward free-falling observer (or test particle) at $r_1>2m$ with
respect to / observed by a stationary observer at $r_2\geq r_1$, according to the results of Section \ref{free-falling}. The radial coordinate that we are going to use is $a=\sqrt{1-\frac{2m}{r}}$, taking values from $0$ (when $r\rightarrow 2m$) to $1$ (when $r\rightarrow +\infty$); so, the radial parameters are $a_1=a\left( r_1\right) $ and $a_2=a\left( r_2\right) $. Another parameter is given by the energy $E>0$ of the free falling test particle. In our study, we are going to consider the modulus of the relative velocities as functions of $a_1$, taking $a_2$ and $E$ as parameters. So, taking into account the definition of $E$, it is clear that $0<a_1\leq a_{1\mathrm{ max}}:=\min\left\{ E,a_2\right\} $. %Note that $E=1$ corresponds to an observer with velocity $v_0=0$ at the limit $r_0=+\infty $.

\subsubsection*{A.1.1 Kinematic relative velocity}

The modulus of the kinematic relative velocity is given by
\[
\left\Vert v_{\mathrm{kin}}\right\Vert =\sqrt{1-\frac{a_1^2}{E^2}}.
\]
Note that $\left\Vert v_{\mathrm{kin}}\right\Vert$ does not depend on $a_2$. It satisfies $0\leq \left\Vert v_{\mathrm{kin}}\right\Vert<1$, it is decreasing with $a_1$ (i.e. increasing with time), and $\lim _{a_1\rightarrow 0}\left\Vert v_{\mathrm{kin}}\right\Vert =1$. Moreover:
\begin{itemize}
\item If $E\leq a_2$, then $\left\Vert v_{\mathrm{kin}}\right\Vert $ takes its minimum at $a_1=a_{1\mathrm{ max}}=E$ and it is $0$.
\item If $E>a_2$, then $\left\Vert v_{\mathrm{kin}}\right\Vert $ takes its minimum at $a_1=a_{1\mathrm{ max}}=a_2$ and it is given by
\begin{equation}
\label{apvkin}
\left\Vert v_{\mathrm{kin}}\right\Vert _{\mathrm{min}}:=\sqrt{1-\frac{a_2^2}{E^2}}.
\end{equation}
We have that $\lim _{E\rightarrow +\infty}\left\Vert v_{\mathrm{kin}}\right\Vert _{\mathrm{min}} = 1$.
\end{itemize}

\subsubsection*{A.1.2 Fermi relative velocity}

The modulus of the Fermi relative velocity is given by
\[
\left\Vert v_{\mathrm{Fermi}}\right\Vert =\frac{a_1}{a^2 _2}\sqrt{1-\frac{a_1^2}{E^2}}.
\]
It satisfies $\lim _{a_1\rightarrow 0}\left\Vert v_{\mathrm{Fermi}}\right\Vert =0$.
Moreover:
\begin{itemize}
\item If $E<\sqrt{2}a_2$, then $\left\Vert v_{\mathrm{Fermi}}\right\Vert $ takes its maximum at $a_1=\frac{E}{\sqrt{2}}$ and it is given by
\[
\left\Vert v_{\mathrm{Fermi}}\right\Vert _{\mathrm{max}}:=\frac{E}{2a_2^2}<\frac{1}{\sqrt{2}a_2}.
\]
It is increasing with $E$, becoming \textit{superluminal} (i.e. $>1$) if, in addition, $E>2a_2^2$. Note that it is only possible if $a_2<\frac{1}{\sqrt{2}}$ (i.e. $r_2<4m$). In this case, $\left\Vert v_{\mathrm{Fermi}}\right\Vert $ is \textit{superluminal} if
\[
\frac{E^2}{2}\left( 1-\sqrt{1-4\frac{a_2^4}{E^2}}\right)<a_1^2<\frac{E^2}{2}\left( 1+\sqrt{1-4\frac{a_2^4}{E^2}}\right).
\]
\item If $E\geq \sqrt{2}a_2$, then $\left\Vert v_{\mathrm{Fermi}}\right\Vert $ is increasing with $a_1$ (i.e. decreasing with time) and takes its maximum at $a_1=a_{1\mathrm{ max}}=a_2$, given by
\begin{equation}
\label{apvfer}
\left\Vert v_{\mathrm{Fermi}}\right\Vert _{\mathrm{max}}:=\frac{1}{a_2}\sqrt{1-\frac{a_2^2}{E^2}}.
\end{equation}
It is increasing with $E$, becoming \textit{superluminal} if $E>\frac{a_2}{\sqrt{1-a_2^2}}$; nevertheless, it is bounded by $\lim _{E\rightarrow +\infty}\left\Vert v_{\mathrm{Fermi}}\right\Vert _{\mathrm{max}} = \frac{1}{a_2}>1$. In this case, $\left\Vert v_{\mathrm{Fermi}}\right\Vert $ is \textit{superluminal} if
\[
a_1^2>\frac{E^2}{2}\left( 1-\sqrt{1-4\frac{a_2^4}{E^2}}\right) .
\]
\end{itemize}
On the other hand,
\begin{itemize}
\item If $E\leq a_2$, then $\left\Vert v_{\mathrm{Fermi}}\right\Vert $ takes its minimum at $a_1=a_{1\mathrm{ max}}=E$ and it is $0$.
\item If $a_2<E<\sqrt{2}a_2$, then $\left\Vert v_{\mathrm{Fermi}}\right\Vert $ has a relative minimum at $a_1=a_{1\mathrm{ max}}=a_2$ and it is given by (\ref{apvfer}). Note that it is \textit{superluminal} if, in addition, $E>\frac{a_2}{\sqrt{1-a_2^2}}$.
\end{itemize}

\subsubsection*{A.1.3 Spectroscopic relative velocity}

The modulus of the spectroscopic relative velocity is given by
\[
\left\Vert v_{\mathrm{spec}}\right\Vert =\frac{\left(
a_{2}^{2}+a_{1}^{2}\right) \sqrt{1-\frac{a_1^2}{E^2}}+\left(
a_{2}^{2}-a_{1}^{2}\right) }{\left( a_{2}^{2}-a_{1}^{2}\right)
\sqrt{1-\frac{a_1^2}{E^2}}+\left( a_{2}^{2}+a_{1}^{2}\right) }.
\]
It satisfies $0\leq \left\Vert v_{\mathrm{spec}}\right\Vert<1$, it is decreasing with $a_1$ (i.e. increasing with time), and $\lim _{a_1\rightarrow 0}\left\Vert v_{\mathrm{spec}}\right\Vert =1$. Moreover:
\begin{itemize}
\item If $E\leq a_2$, then $\left\Vert v_{\mathrm{spec}}\right\Vert $ takes its minimum at $a_1=a_{1\mathrm{ max}}=E$ and it is given by
\[
\left\Vert v_{\mathrm{spec}}\right\Vert _{\mathrm{min}}:=\frac{a_2^2-E^2}{a_2^2+E^2}.
\]
We have that $\left\Vert v_{\mathrm{spec}}\right\Vert _{\mathrm{min}}$ is decreasing with $E$, and it only vanishes at $E=a_2$.
\item If $E>a_2$, then $\left\Vert v_{\mathrm{spec}}\right\Vert $ takes its minimum at $a_1=a_{1\mathrm{ max}}=a_2$ and it is given by
\[
\left\Vert v_{\mathrm{spec}}\right\Vert _{\mathrm{min}}:=\sqrt{1-\frac{a_2^2}{E^2}}.
\]
Note that this is the same minimum as in the kinematic case (see (\ref{apvkin})).
\end{itemize}

\subsubsection*{A.1.4 Astrometric relative velocity}

The modulus of the astrometric relative velocity is given by
\[
\left\Vert v_{\mathrm{ast}}\right\Vert
=\frac{a_{1}^{2}}{a^2_{2}}\frac{\sqrt{1-\frac{a_1^2}{E^2}}}{1+\left(
a_{1}^{2}-1\right) \sqrt{1-\frac{a_1^2}{E^2}}}.
\]
It is important to note that $\lim _{E\rightarrow +\infty }\left\Vert v_{\mathrm{ast}}\right\Vert = \frac{1}{a_2^2}>1$ for all $a_1$. So, given $a_2$, there exists always a big enough energy (see (\ref{apvast}) below) such that $\left\Vert v_{\mathrm{ast}}\right\Vert$ is \textit{superluminal} for all $a_1$.

It is decreasing with $a_1$ (i.e. increasing with time), and it has a supremum
\[
\left\Vert v_{\mathrm{ast}}\right\Vert _{\mathrm{sup}}:=\lim_{a_{1}\rightarrow 0}\left\Vert
v_{\mathrm{ast}}\right\Vert =\frac{1}{a_{2}^{2}}\frac{2E^{2}}{1+2E^{2}}.
\]
We have that $\left\Vert v_{\mathrm{ast}}\right\Vert _{\mathrm{sup}}$ is increasing with $E$, becoming \textit{superluminal} if $E>\frac{1}{\sqrt{2}}\frac{a_2}{\sqrt{1-a_2^2}}$ (but it is bounded by $\frac{1}{a_2^2}$). In this case, $\left\Vert v_{\mathrm{ast}}\right\Vert $ is \textit{superluminal} if
\[
a_1^2<\frac{E^2}{2}\left( 1+\sqrt{1+\frac{4}{E^2}\frac{a_2^2}{1-a_2^2}}\right)-\frac{a_2^2}{1-a_2^2}.
\]
Moreover:
\begin{itemize}
\item If $E\leq a_2$, then $\left\Vert v_{\mathrm{ast}}\right\Vert $ takes its minimum at $a_1=a_{1\mathrm{ max}}=E$ and it is $0$.
\item If $E>a_2$, then $\left\Vert v_{\mathrm{ast}}\right\Vert $ takes its minimum at $a_1=a_{1\mathrm{ max}}=a_2$ and it is given by
\[
\left\Vert v_{\mathrm{ast}}\right\Vert _{\mathrm{min}}:=\frac{\sqrt{1-\frac{a_2^2}{E^2}}}{1+\left(
a_2^2-1\right) \sqrt{1-\frac{a_2^2}{E^2}}}.
\]
It is increasing with $E$, becoming \textit{superluminal} if
\begin{equation}
\label{apvast}
E>\frac{a_2\left( 2-a_2^2\right) }{\sqrt{\left( 2-a_2^2\right) ^2-1}}.
\end{equation}
\end{itemize}

\begin{figure}[tbp]
\begin{center}
\includegraphics[width=1\textwidth]{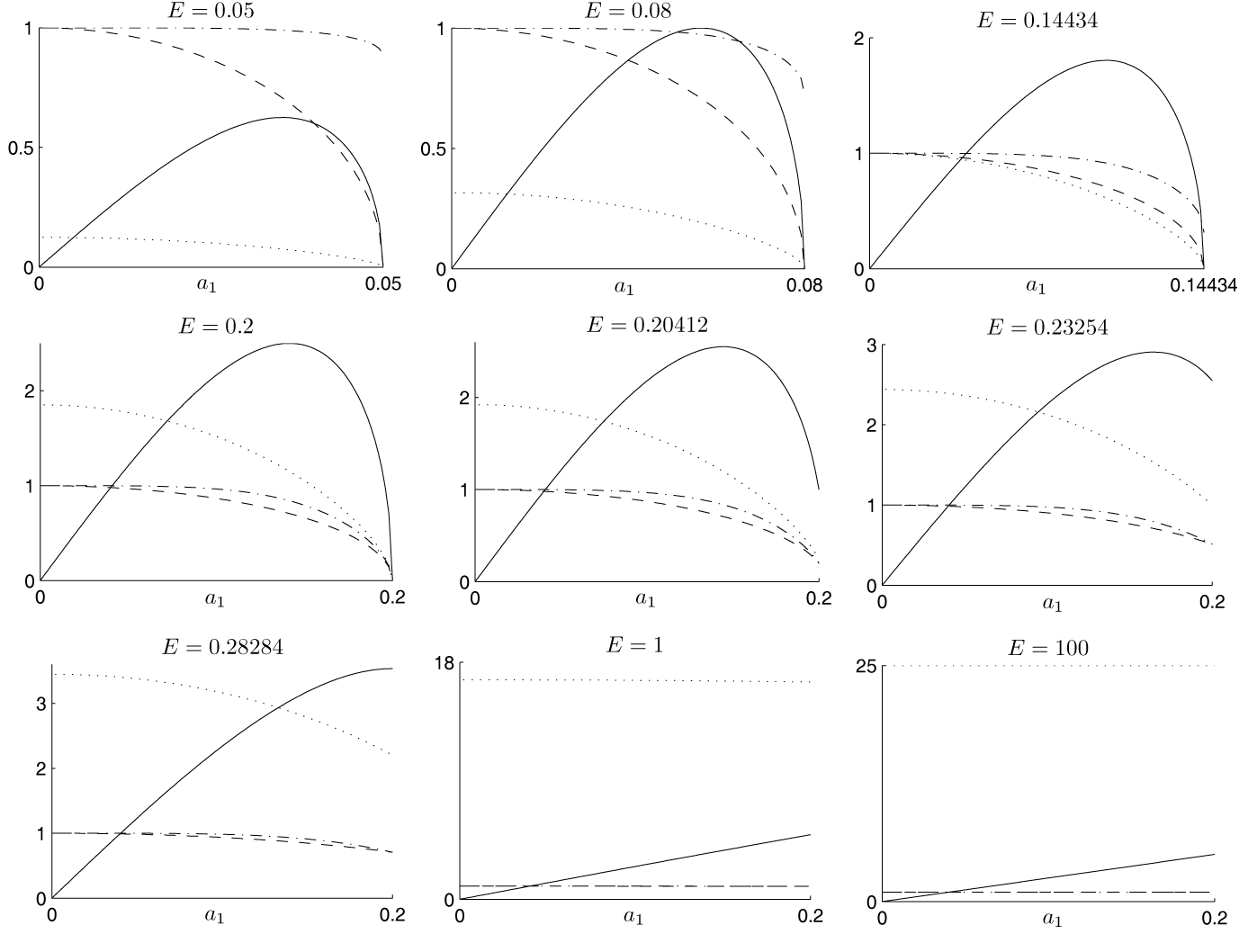}
\end{center}
\caption{Moduli of kinematic (dashed), Fermi (solid), spectroscopic (dot-dashed) and astrometric (dotted) relative velocities with $a_2=0.2$. At $E=0.08$ (top center), $\left\Vert v_{\mathrm{Fermi}}\right\Vert _{\mathrm{max}}$ begins to be \textit{superluminal}. At $E=0.14434$ (top right), $\left\Vert v_{\mathrm{ast}}\right\Vert _{\mathrm{sup}}$ begins to be \textit{superluminal}. At $E=a_2=0.2$ (middle left), all the velocities vanish at $a_1=a_{1\mathrm{ max}}=0.2$, and these minima begin to increase for higher energies. At $E=0.20412$ (middle center), the relative minimum of $\left\Vert v_{\mathrm{Fermi}}\right\Vert $ at $a_1=0.2$ begins to be \textit{superluminal}. At $E=0.23254$ (middle right), $\left\Vert v_{\mathrm{ast}}\right\Vert _{\mathrm{min}}$ begins to be \textit{superluminal}. At $E=0.28284$ (bottom left), $\left\Vert v_{\mathrm{Fermi}}\right\Vert $ begins to be monotonic.}
\label{ff_a2-020}
\end{figure}

\begin{figure}[tbp]
\begin{center}
\includegraphics[width=1\textwidth]{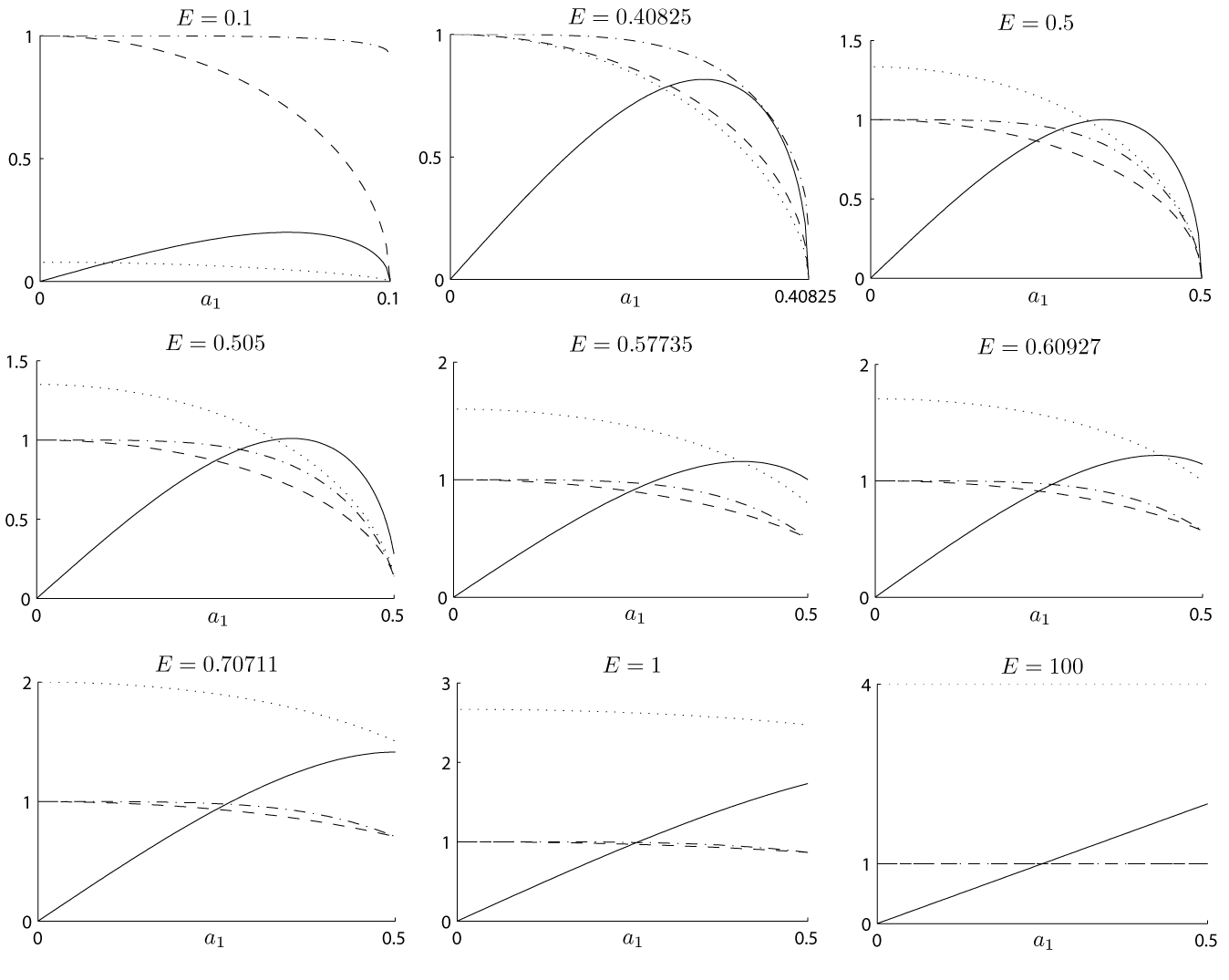}
\end{center}
\caption{Moduli of kinematic (dashed), Fermi (solid), spectroscopic (dot-dashed) and astrometric (dotted) relative velocities with $a_2=0.5$. At $E=0.40825$ (top center), $\left\Vert v_{\mathrm{ast}}\right\Vert _{\mathrm{sup}}$ begins to be \textit{superluminal}. At $E=a_2=0.5$ (top right), all the velocities vanish at $a_1=a_{1\mathrm{ max}}=0.5$, and these minima begin to increase for higher energies; moreover $\left\Vert v_{\mathrm{Fermi}}\right\Vert _{\mathrm{max}}$ begins to be \textit{superluminal}. At $E=0.57735$ (middle center), the relative minimum of $\left\Vert v_{\mathrm{Fermi}}\right\Vert $ at $a_1=0.5$ begins to be \textit{superluminal}. At $E=0.60927$ (middle right), $\left\Vert v_{\mathrm{ast}}\right\Vert _{\mathrm{min}}$ begins to be \textit{superluminal}. At $E=0.70711$ (bottom left), $\left\Vert v_{\mathrm{Fermi}}\right\Vert $ begins to be monotonic.}
\label{ff_a2-050}
\end{figure}

\begin{figure}[tbp]
\begin{center}
\includegraphics[width=1\textwidth]{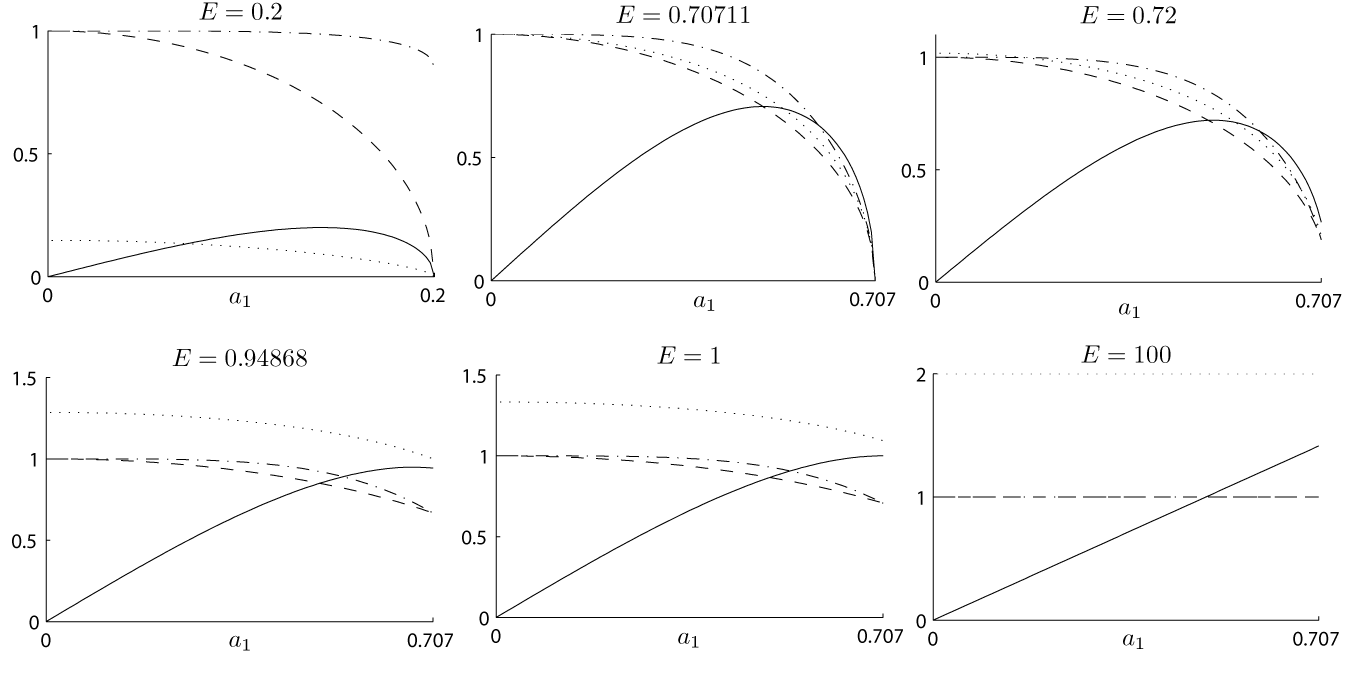}
\end{center}
\caption{Moduli of kinematic (dashed), Fermi (solid), spectroscopic (dot-dashed) and astrometric (dotted) relative velocities with $a_2=0.70711$ $(r_2=4m)$. At $E=a_2=0.70711$ (top center), all the velocities vanish at $a_1=a_{1\mathrm{ max}}=0.70711$, and these minima begin to increase for higher energies; moreover $\left\Vert v_{\mathrm{ast}}\right\Vert _{\mathrm{sup}}$ begins to be \textit{superluminal}. At $E=0.94868$ (bottom left), $\left\Vert v_{\mathrm{ast}}\right\Vert _{\mathrm{min}}$ begins to be \textit{superluminal}. At $E=1$ (bottom center), $\left\Vert v_{\mathrm{Fermi}}\right\Vert $ begins to be monotonic and its maximum at $a_1=0.70711$ begins to be \textit{superluminal}.}
\label{ff_r2-4m}
\end{figure}

\begin{figure}[tbp]
\begin{center}
\includegraphics[width=1\textwidth]{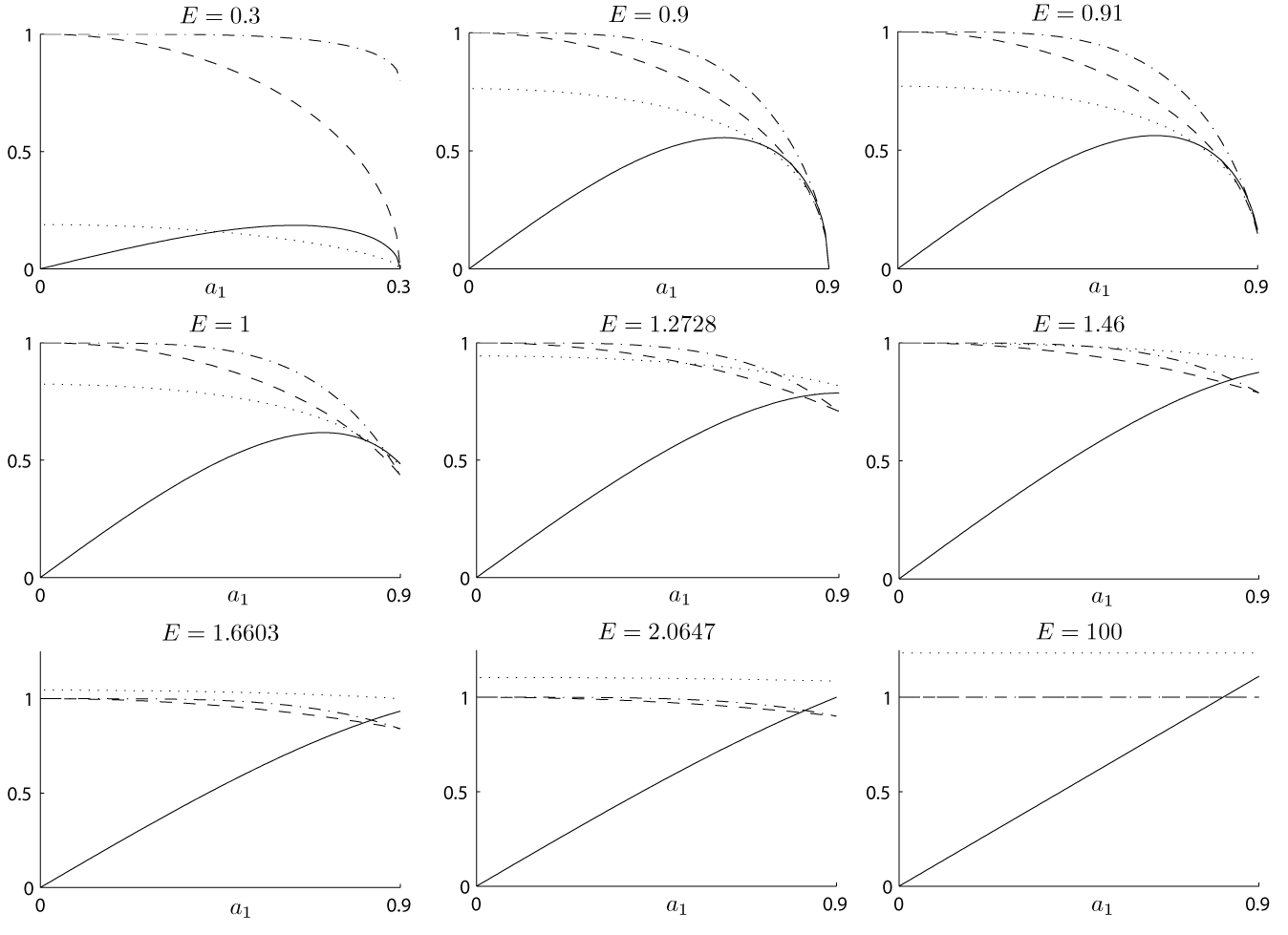}
\end{center}
\caption{Moduli of kinematic (dashed), Fermi (solid), spectroscopic (dot-dashed) and astrometric (dotted) relative velocities with $a_2=0.9$. At $E=a_2=0.9$ (top center), all the velocities vanish at $a_1=a_{1\mathrm{ max}}=0.9$, and these minima begin to increase for higher energies. At $E=1.2728$ (middle center), $\left\Vert v_{\mathrm{Fermi}}\right\Vert $ begins to be monotonic. At $E=1.46$ (middle right), $\left\Vert v_{\mathrm{ast}}\right\Vert _{\mathrm{sup}}$ begins to be \textit{superluminal}. At $E=1.6603$ (bottom left), $\left\Vert v_{\mathrm{ast}}\right\Vert _{\mathrm{min}}$ begins to be \textit{superluminal}. At $E=2.0647$ (bottom center), $\left\Vert v_{\mathrm{Fermi}}\right\Vert _{\mathrm{max}}$ begins to be \textit{superluminal}.}
\label{ff_a2-090}
\end{figure}

\begin{figure}[tbp]
\begin{center}
\includegraphics[width=0.95\textwidth]{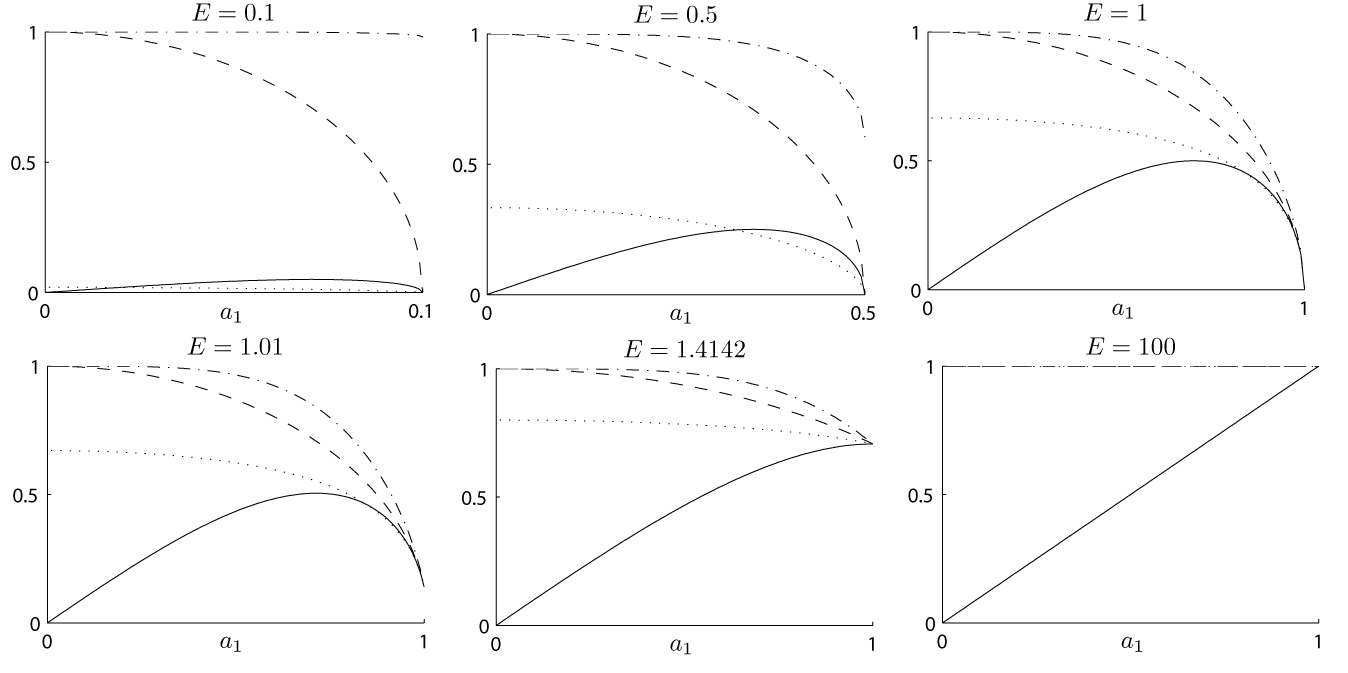}
\end{center}
\caption{Moduli of kinematic (dashed), Fermi (solid), spectroscopic (dot-dashed) and astrometric (dotted) relative velocities in the exterior limit $a_2=1$. There is not any \textit{superluminal} velocity. At $E=a_2=1$ (top right), all the velocities vanish at $a_1=a_{1\mathrm{ max}}=1$, and this minimum (note that all the velocities have the same minimum) begins to increase for higher energies. At $E=\sqrt{2}$ (bottom center), $\left\Vert v_{\mathrm{Fermi}}\right\Vert $ begins to be monotonic.}
\label{ff_a2-100}
\end{figure}

See Figures \ref{ff_a2-020} ($a_2=0.2$), \ref{ff_a2-050} ($a_2=0.5$), \ref{ff_r2-4m}, ($a_2=0.70711$, i.e. $r_2=4m$), \ref{ff_a2-090}  ($a_2=0.9$), and \ref{ff_a2-100} (exterior limit $a_2=1$). In all figures at low energies (top left) there is not any \textit{superluminal} velocity and all the velocities vanishes at $a_1=a_{1\mathrm{ max}}=E$ except for $\left\Vert v_{\mathrm{spec}}\right\Vert $. At $E=a_2$, all the velocities vanish at $a_1=a_{1\mathrm{ max}}=E=a_2$, and these minima begin to increase for higher energies; moreover, $\left\Vert v_{\mathrm{kin}}\right\Vert $ and $\left\Vert v_{\mathrm{spec}}\right\Vert $ have the same minimum. At high energies (bottom right), $\left\Vert v_{\mathrm{kin}}\right\Vert $ and $\left\Vert v_{\mathrm{spec}}\right\Vert $ tends to $1$, $\left\Vert v_{\mathrm{Fermi}}\right\Vert $ tends to $\frac{a_1}{a_2^2}$, and $\left\Vert v_{\mathrm{ast}}\right\Vert $ tends to $\frac{1}{a_2^2}$.

\section*{Acknowledgments}

I would like to thank Ettore Minguzzi, Pedro Sancho, Vicente Miquel, David Klein and the referees of the journal \textit{Communications in Mathematical Physics} for their valuable
help and comments.

\end{document}